\documentclass[preprint,aps,12pt,showpacs,nofootinbib,tightenlines]{revtex4}
\usepackage{mathrsfs}
\usepackage{amsmath}
\usepackage{amssymb}
\usepackage{epsfig}
\usepackage{epstopdf}
\usepackage{graphicx}
\usepackage{subfigure}
\usepackage{booktabs}
\usepackage{float}
\usepackage{amsmath}
\usepackage{color}
\usepackage{multirow}
\usepackage{bm}

\def\be{\begin{eqnarray}}
\def\en{\end{eqnarray}}
\def\non{\nonumber\\}

\begin{document}
\title{$D_{(s)}(2S)$ and $D^{*}_{(s)}(2S)$ production in nonleptonic $B_{(s)}$ weak decays}
\author{Zhi-Jie Sun$^1$, 
Yong-Jin Sun$^1$,
Zhi-Qing Zhang$^2$\footnote{zhangzhiqing@haut.edu.cn}
You-Ya Yang$^3$, Si-Yang Wang$^4$} 
\affiliation{\it \small $^1$ Bingtuan Xingxin Vocational and Technical College, Tiemenguan, Xinjiang, 841007, China\\
\it \small $^2$ School of Physics and advanced energy, Henan University of Technology, Zhengzhou, Henan 450001, China \\
\it \small $^3$ Physics Department, College of Physics and Optoelectronic Engineering, Jinan University, Guangzhou 510632, China\\
\it \small $^4$ Huanghe Jiaotong University, Jiaozuo, Henan 454950, China }
\date{\today}
\begin{abstract}
Recently, many new excited states of heavy mesons have been discovered in recent experiments, including radially excited states. The production processes of these states from the $B_{(s)}$ meson have drawn significant interest. In this paper, we use the covariant light-front approach to study the nonleptonic $B_{(s)}$ meson decays to the first radially excited states $D_{(s)}(2S)$ and $D^{*}_{(s)}(2S)$. Our results reveal that many channels exhibit large branching ratios in the range $10^{-5}\sim 10^{-4}$, even up to $10^{-3}$ for individual channels, which are detectable by current experiments. Our predictions for the decays $B_{(s)}\to D^{(*)}_{(s)}(2S)(\pi,\rho,K^{(*)})$ are larger than those given by the Bethe-Salpeter (BS) equation method, but agree well with the relativistic quark mode (RQM) and the relativistic independent quark model (RIQM) calculations.  
For comparison, we also present the branching ratios of the decays $B_{(s)}\to D^{(*)}_{(s)}(1S)(\pi,\rho,K^{(*)})$, which are comparable with other theoretical results and the data. Although the branching ratios of the decays $B_{(s)} \to D^{*}_{(s)}(1S)(\rho,K^*)$ are much larger than those of the decays $B_{(s)} \to D^{*}_{(s)}(2S)(\rho,K^*)$, the polarization properties between them are similar, that is, the longitudinal polarization fractions are dominant and can amount roughly to $90\%$. 
\end{abstract}

 \vspace{1cm}

\maketitle

\section{Introduction}\label{intro}
 A wealth of experimental data on $B_{(s)}$ meson decays have been collected, allowing studies of both ground and radially excited final states. Due to their heavy masses, $B_{(s)}$ meson exhibit rich decay modes \cite{PDGT2024}. These decays offer an opportunity to search for the radially excited states and clarify their properties. For example, $\eta_c(2S)$ was discovered in exclusive decays $B\to KK_SK^-\pi^+$ \cite{belle} more than two decades ago. The well-determined radially excited charmonium states, such as $\psi(2S)$ and $\eta_c(2S)$, exhibit a significant production yield in the nonleptonic decays of
$B$ meson. One can find that many ratios 
$\frac{Br(B\to \psi(2S)M)}{Br(B^+\to J/\Psi M)}$ with $M$ being a light pseudoscalar or vector meson are around or even larger than $50\%$ \cite{PDGT2024}. What happens if charmonium states are replaced by charmed mesons?
Although many first radially excited (2S) state candidates of the ground charmed mesons $D^{(*)}$ and $D^{(*)}_s$ have been found in experiments, they are still not well-determined. Usually, $D(2550)^0$ is classified as 2S state $2^1S_0$ in the D meson family. It was first observed by BaBar in the $e^+e^-\to D^{*+}\pi^-$ channel in 2010 \cite{BaBar}, and was confirmed by LHCb with significance  using pp collision data. The assignment of $D(2^1S_0)$ to $D(2550)^0$ was supported by various quark models \cite{liujb,bada,song}. In addition, another unnatural state $D_J(2580)^0$ found by LHCb may also be a candidate for $D(2^1S_0)$, since $D_J(2580)^0$ and $D(2550)^0$ have similar resonance parameters. 
$D_{s0}(2590)^+$ with $J^P=0^-$ newly observed by LHCb in the $D^+K^+\pi^-$ invariant mass spectrum of the decay $B^0\to D^+D^+K^+\pi^-$ is suggested for a candidate of $D_s(2^1S_0)$ state \cite{lhcb2590}. Certainly, this suggestion has difficulties accounting for the mass
and width of $D_{s0}(2590)^+$. Some theoretical works suggest that $D_{s0}(2590)^+$ may not be a pure $D_s(2^1S_0)$ state but rather has $D^*K$ component \cite{xie,ortega}. Regarding the neutral radially excited 2S state $D(2^3S_1)$, there are several candidates, such as $D_{1}^{*}(2680)^{0}, D^{*}(2650)^{0}$ and $D^*_1(2600)^0$ observed by LHCb \cite{LHCb2016,LHCb2013,LHCb2019}. There is still some uncertainty in the measurements, and whether they are identical cannot be determined, but the spin-parity of $D_{1}^{*}(2680)^{0}$ and  $D^*_1(2600)^0$ can be confirmed as $1^-$. 
The mass of $D^*(2640)^\pm$ discovered by Delphi is consistent with the prediction of the charged state $D(2^3S_1)$ \cite{godfrey1}. Unfortunately, this discovery has not been confirmed in any other experiments and its spin-parity numbers have not been identified up to now.
The $D^{*}_{s1}(2700)^{\pm}$ can be assigned to the state $D_s(2^3S_1)$, which was first observed by BaBar \cite{aubert6} and whose quantum numbers $J^P$ were determined as $1^{-}$ by Belle \cite{belle8}. Certainly, except for the assignment of $D_s(2^3S_1)$ to $D^{*}_{s1}(2700)^{\pm}$,  $D_{s}(1^3D_1)$ \cite{godfrey1} and a mixture of $D_s(2^3S_1)$ and $D_{s}(1^3D_1)$ \cite{Ebert} have also been proposed. In a word, we will assume the $D(2550)^0$ as neutral state $D(2^1S_0)$, the $D_{s0}(2590)^\pm$ as $D_s(2^1S_0)$, the $D^*_1(2600)^0, D^*(2640)^\pm$ as  $D(2^3S_1)$, the $D^*_{s1}(2700)^\pm$ as $D_s(2^3S_1)$, respectively. It is noticed that no candidate for the charged $D(2^1S_0)$ state has been observed in experiments up to now.  

For the nonleptonic decays, the hadronic transition matrix element between the initial and final mesons is most crucial for the theoretical calculations. The factorization
 assumption based on the vacuum saturation approximation is often used to simplify the calculations. Specifically, the matrix elements are factorized into a product of two single matrix elements of currents, where one is parameterized by the decay constant of the emitted light meson and the other is represented by
the transition form factor. As for the form factors, they can be extracted from data or relied on some nonperturbative and perturbative methods, such as the Bauer-Stech-Wirbel (BSW) model \cite{R.R},  the QCD light-cone sum rules (LCSR)  \cite{Wu:2025kdc,Fu:2013wqa,Zhang:2017rwz,Gao:2021sav,Zhong:2018exo,Li:2009wq}, the lattice QCD (LQCD) \cite{Yao:2019vty,MILC:2015uhg,Na:2015kha}, the relativistic quark model (RQM) \cite{Faustov:2022ybm,Faustov:2012mt}, the QCD sum rules (QCDSR) \cite{YMW,YMW2}, the Bethe-Salpeter (BS) method \cite{Wang:2016dkd}, the covariant light front quark model (CLFQM) \cite{Zhang:2023ypl} and the perturbative QCD (PQCD) approach \cite{lirh}.

This paper is organized as follows. The formalism of the CLFQM, the hadronic matrix elements and the helicity amplitudes combined via form factors are listed in Sec. \ref{form1}. In addition to the numerical results for the
$B_{(s)}\to D_{(s)}(1S,2S)$ and $B_{(s)}\to D^{*}_{(s)}(1S,2S)$ transition form factors, the branching ratios, the longitudinal (transverse parallel) polarization fractions $f_{L}(f_{\|})$ of the corresponding decays are presented in Sec. \ref{numer}. Detailed comparisons with other theoretical values and relevant discussions are also included. The summary is presented in Sec. \ref{sum}. Some specific rules when performing the $p^-$ integration and the analytical expressions for the $B_{(s)}\to D_{(s)}(1S,2S)$ and $B_{(s)}\to D^{*}_{(s)}(1S,2S)$ transition form factors are collected in Appendixes A and B, respectively.
\section{Formalism}\label{form1}
\subsection{The form factors}
The Bauer-Stech-Wirbel (BSW) form factors for the $B_{(s)} \rightarrow D_{(s)}$ and $B_{(s)} \rightarrow D^{*}_{(s)}$ transitions are defined as follows,
\begin{footnotesize}
\begin{eqnarray}
\left\langle D_{(s)}\left(P^{\prime
\prime}\right)\left|V_{\mu}\right|
B_{(s)}\left(P^{\prime}\right)\right\rangle
&=&\left(P_{\mu}-\frac{m_{B_{(s)}}^{2}-m_{D_{(s)}}^{2}}{q^{2}}
q_{\mu}\right) F_{1}^{B_{(s)}
D_{(s)}}\left(q^{2}\right)+\frac{m_{B_{(s)}}^{2}-m_{D_{(s)}}^{2}}{q^{2}} q_{\mu}\nonumber \\
&&\cdot F_{0}^{B_{(s)} D_{(s)}}\left(q^{2}\right),\\    
\label{pdep} 
\left\langle D^{*}_{(s)}\left(P^{\prime \prime},\epsilon^{\prime\prime*}\right)\left|V_{\mu}-A_{\mu}\right| B_{(s)} \left(P^{\prime}\right)\right\rangle
&=&-\epsilon_{\mu \nu \alpha \beta} \epsilon^{\prime\prime}_\mu P^{\alpha} q^{\beta} \frac{V\left(q^{2}\right)}{m_{B_{(s)}}+m_{D^{*}_{(s)}}}-i \frac{2 m_{D^*_{(s)}} \epsilon^{\prime\prime*}\cdot P}{q^{2}} q_{\mu} A_{0}\left(q^{2}\right) \nonumber \\
&&-i \epsilon^{\prime\prime*}_{\mu}\left(m_{B_{(s)}}+m_{D^{*}_{(s)}}\right) A_{1}\left(q^{2}\right)+i \frac{\epsilon^{\prime\prime*} \cdot P}{m_{ B_{(s)}}+m_{D^{*}_{(s)}}}P_{\mu} A_{2}\left(q^{2}\right) \nonumber \\
&&+i \frac{2 m_{D^{*}_{(s)}} \epsilon^{\prime\prime*} \cdot P}{q^{2}} q_{\mu} A_{3}\left(q^{2}\right),\label{pdev}
\end{eqnarray}
\end{footnotesize}

where $P=P'+P'', q=P'-P''$ and the convention $\epsilon_{0123}=1$ is adopted. 

\begin{figure}[htbp]
\centering \subfigure{
\begin{minipage}{5cm}
\centering
\includegraphics[width=5cm]{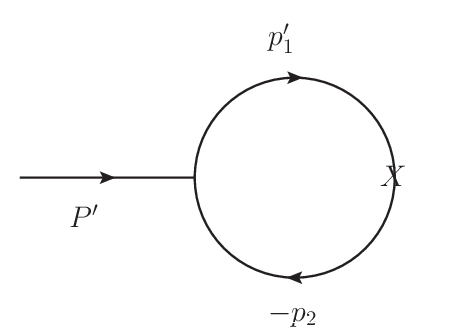}
\end{minipage}}
\subfigure{
\begin{minipage}{6cm}
\centering
\includegraphics[width=6cm]{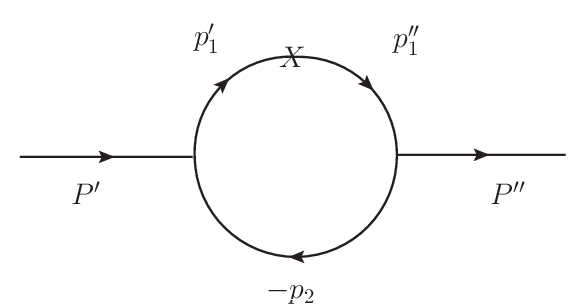}
\end{minipage}}
\caption{Feynman diagrams for $B_{(s)}$ decay (left) and transition
(right) amplitudes, where $P^{\prime(\prime\prime)}$ is the
incoming (outgoing) meson momentum, $p^{\prime(\prime\prime)}_1$
is the quark momentum, $p_2$ is the anti-quark momentum and X
denotes the vector or axial-vector transition vertex.}
\label{feyn}
\end{figure}

In order to calculate the amplitudes of the transition form factors, we need the following Feynman rules for the meson-quark-antiquark vertices
\be
i\Gamma^{\prime} _{P}&=&H^{\prime}_{P}\gamma_{5},\\
i\Gamma^{\prime\prime} _{P}&=&\gamma_0H^{\prime\prime}_{P}\gamma_{5}\gamma_0,\\
i\Gamma^{\prime\prime} _{V}&=&i \gamma_0H_{V}^{\prime\prime}\left[\gamma_{\mu}-\frac{1}{W_{V}^{\prime}}\left(p_{1}^{\prime}-p_{2}\right)_{\mu}\right]\gamma_0,
\en
where the last two formulas are for the final state mesons. 
The results of the lowest-order transition form factors could be obtained by calculating the right Feynman diagram in Figure \ref{feyn}, where the left panel is for the decay amplitude of $B_{(s)}$ meson. In the covariant quark model, the treatment of transition form factors is relatively covariant throughout the calculation process, where the light-front coordinates of a momentum $p$ are used $p=(p^-,p^+,p_\perp)$ with
$p^\pm=p^0\pm p_z, p^2=p^+p^--p^2_\perp$.
The incoming (outgoing) meson has the mass $M^\prime(M^{\prime\prime})$
with the momentum $P^\prime=p_1^\prime+p_2 (P^{\prime\prime}=p_1^{\prime\prime}+p_2)$, where $p_{1}^{\prime(\prime\prime)} $
and $p_{2}$ are the momenta of the quark and anti-quark
inside the incoming (outgoing) meson with the mass $m_{1}^{\prime(\prime\prime)}$and $m_{2}$, respectively. Here we use the same notations
as those in Refs. \cite{jaus,Y. Cheng}.
These momenta can be expressed in terms of the internal variables $(x_{i},p{'}_{\perp})$ as
\be
p_{1,2}^{\prime+}=x_{1,2} P^{\prime+}, \quad p_{1,2 \perp}^{\prime}=x_{1,2} P_{\perp}^{\prime} \pm p_{\perp}^{\prime}
\en
with $x_{1}+x_{2}=1$. Using these internal variables,
we can define some quantities for the incoming meson which will be used in the following calculations
\be
M_{0}^{\prime 2} &=&\left(e_{1}^{\prime}+e_{2}\right)^{2}=\frac{p_{\perp}^{\prime 2}+m_{1}^{\prime 2}}{x_{1}}
+\frac{p_{\perp}^{2}+m_{2}^{2}}{x_{2}}, \quad \widetilde{M}_{0}^{\prime}=\sqrt{M_{0}^{\prime 2}-\left(m_{1}^{\prime}-m_{2}\right)^{2}},\non
e_{i}^{(\prime)} &=&\sqrt{m_{i}^{(\prime) 2}+p_{\perp}^{\prime 2}+p_{z}^{\prime 2}}, \quad \quad p_{z}^{\prime}
=\frac{x_{2} M_{0}^{\prime}}{2}-\frac{m_{2}^{2}+p_{\perp}^{\prime 2}}{2 x_{2} M_{0}^{\prime}},
\en
where $M'_0$ is the kinetic invariant mass of the incoming meson and can be expressed as the energies of the quark and the anti-quark
$e^{(\prime)}_i$. It is similar to the case of the outgoing meson.
For the general $P\to P$ transition, the amplitude for the lowest order is
\be
\mathcal{B}_{\mu}^{P P}=-i^{3} \frac{N_{c}}{(2 \pi)^{4}} \int d^{4} p_{1}^{\prime} \frac{H_{P}^{\prime}H_{P}^{\prime\prime}}
{N_{1}^{\prime} N_{1}^{\prime \prime} N_{2}} S_{\mu}^{P P},
\label{ZF}
\en
where $N_{1}^{\prime(\prime \prime)}=p_{1}^{\prime(\prime \prime) 2}-m_{1}^{\prime (\prime\prime) 2}, N_{2}=p_{2}^{2}-m_{2}^{2} $ arise
from the quark propagators, and
the trace $S_{\mu}^{PP}$ can be obtained directly by using the Lorentz contraction,
\be
S_{\mu}^{P P}&=&\operatorname{Tr}\left[\gamma_{5}\left(\not p_{1}^{\prime \prime}+m_{1}^{\prime \prime}\right) \gamma_{\mu}\left(\not p_{1}^{\prime}
+m_{1}^{\prime}\right) \gamma_{5}\left(-\not p_{2}+m_{2}\right)\right],
\label{ptop}
\en
where the analytical expression for $S_{\mu}^{P P}$ is listed in Appendix B.  It is similar for the $P\to V$ transition amplitude,
\be
\mathcal{B}_{\mu}^{P V}=-i^{3} \frac{N_{c}}{(2 \pi)^{4}} \int d^{4} p_{1}^{\prime} \frac{H_{P}^{\prime}\left(i H_{V}^{\prime \prime}\right)}{N_{1}^{\prime} N_{1}^{\prime \prime} N_{2}}
 S_{\mu \nu}^{P V} \varepsilon^{*\nu},
\en
where
\be
S_{\mu \nu}^{P V}&=&\left(S_{V}^{P V}-S_{A}^{P V}\right)_{\mu \nu}\non
&=&\operatorname{Tr}\left[\left(\gamma_{\nu}-\frac{1}{W_{V}^{\prime \prime}}\left(p_{1}^{\prime \prime}-p_{2}\right)_{\nu}\right)\left(p_{1}^{\prime \prime}
+m_{1}^{\prime \prime}\right)\left(\gamma_{\mu}-\gamma_{\mu} \gamma_{5}\right)\left(\not p_{1}^{\prime}+m_{1}^{\prime}\right) \gamma_{5}\left(-\not p_{2}
+m_{2}\right)\right].\;\;\;
\label{sptov}
\en

In practice, we use the light-front decomposition of the Feynman loop momentum and integrate out
the minus component through the contour method. If the covariant vertex functions are not singular when performing integration,
the transition amplitudes will
pick up the singularities in the anti-quark propagators. The integration then leads to
\be
N_{1}^{\prime(\prime \prime)} &\rightarrow& \hat{N}_{1}^{\prime(\prime \prime)}=x_{1}\left(M^{\prime(\prime \prime) 2}-M_{0}^{\prime(\prime \prime) 2}\right),\non
H_{M}^{\prime(\prime\prime)} &\rightarrow& h_{M}^{\prime(\prime \prime)}\non
W_{V}^{\prime \prime} &\rightarrow& w_{V}^{\prime \prime} \non
\int \frac{d^{4} p_{1}^{\prime}}{N_{1}^{\prime} N_{1}^{\prime \prime} N_{2}} H_{P}^{\prime} H_{M}^{\prime \prime} S^{PM} & \rightarrow&-i \pi \int \frac{d x_{2} d^{2}
p_{\perp}^{\prime}}{x_{2} \hat{N}_{1}^{\prime} \hat{N}_{1}^{\prime \prime}} h_{P}^{\prime} h_{M}^{\prime \prime} \hat{S}^{PM},
\en
where
\be
M_{0}^{\prime \prime 2}=\frac{p_{\perp}^{\prime \prime 2}+m_{1}^{\prime \prime 2}}{x_{1}}+\frac{p_{\perp}^{\prime \prime 2}+m_{2}^{2}}{x_{2}},
\label{vertex}
\en
with $p''_\perp=p'_\perp-x_2q_\perp$. The explicit forms of $h^{\prime\prime}_{M}$  and $w^{\prime\prime}_{V}$ are given by \cite{Y. Cheng}
\be
h_{P}^{\prime\prime} &=&h_{V}^{\prime\prime}=\left(M^{\prime\prime 2}-M_{0}^{\prime\prime 2}\right) \sqrt{\frac{x_{1} x_{2}}{N_{c}}} \frac{1}{\sqrt{2} \widetilde{M}_{0}^{\prime}} \varphi^{\prime\prime},\\
w^{\prime\prime}_{V}&=&M^{\prime\prime}_{0}+m^{\prime\prime}_{1}+m_{2},\label{hp}
\en
where $\varphi^{\prime\prime}$ is the light-front momentum distribution amplitude for S-wave mesons,
\be
\varphi^{\prime\prime} &=&\varphi^{\prime\prime}\left(x_{2}, p_{\perp}^{\prime\prime}\right)=4\left(\frac{\pi}{\beta^{2}}\right)^{\frac{3}{4}}
\sqrt{\frac{d p_{z}^{\prime\prime}}{d x_{2}}} \exp \left(-\frac{p_{z}^{\prime\prime 2}+p_{\perp}^{\prime\prime 2}}{2 \beta^{2}}\right).
\en
where $\beta$ is a phenomenological parameter and can be fixed by fitting the corresponding decay constant. Regarding the first radially excited charmed mesons $D_{(s)}^{*}(2S)$ and $D_{(s)}(2S)$, the distribution function is given as
\be
\varphi^{\prime\prime}(2S) &=&4\left(\frac{\pi}{\beta^{2}}\right)^{\frac{3}{4}}
\sqrt{\frac{d p_{z}^{\prime\prime}}{d x_{2}}} \exp \left(-\frac{p_{z}^{\prime\prime 2}+p_{\perp}^{\prime\prime 2}}{2 \beta^{2}}\right)
\times \frac{1}{\sqrt{6}}\left(-3+2\frac{p_{z}^{\prime\prime 2}+p_{\perp}^{\prime\prime 2}}{\beta^{2}}\right).
\en
Using the formulas provided above and taking the integration rules given in Refs \cite{jaus,Y. Cheng},
we obtain the expressions of the $B_{(s)}\to D_{(s)}(1S,2S)$ and $B_{(s)}\to D^*_{(s)}(1S,2S)$ transition form factors, which are listed in Appendix B.
\subsection{Hadronic matrix elements}
We present the formulas for the nonleptonic decays $B\to DM$ and $B\to D^*M$\footnote{It is similar for the decays $B_s\to D_sM$ and $B_s\to D^*_sM$.} with $M=\pi, K, \rho, K^*$. The effective Hamiltonian which describes such processes can be written as \cite{hep-ph/9705292}
\be
\mathcal{H}_{\mathrm{eff}}=\frac{G_{F}}{\sqrt{2}}  V_{c b}^{*} V_{u q} \left\{C_{1} Q_{1}+C_{2} Q_{2}\right\}
\en
where $G_{F}$ is the Fermi coupling constant, $V^{*}_{cb}V_{uq}$ is the product of the CKM matrix elements with $q=s, d$, and $C_{1,2}$ are the Wilson coefficients.
The local tree-level four-quark operators $Q_{1,2}$ are defined as
\be
Q_{1}=\left[\bar{c}_{\alpha} \gamma_{\mu}\left(1-\gamma_{5}\right) b_{\beta}\right]\left[\bar{q}_{\beta} \gamma^{\mu}\left(1-\gamma_{5}\right) u_{\alpha}\right], \\
Q_{2}=\left[\bar{c}_{\alpha} \gamma_{\mu}\left(1-\gamma_{5}\right) b_{\alpha}\right]\left[\bar{q}_{\beta} \gamma^{\mu}\left(1-\gamma_{5}\right) u_{\beta}\right],
\en
where $\alpha$ and $\beta$ are color indices. In the following  calculations, the combination of Wilson coefficients $a_1=C_2+C_1/3$ will be used. Using these form factors, we can obtain the partial widths for our considered nonleptonic decays, which are written as
\be
\Gamma\left(B \rightarrow DP\right) & =&\frac{G_{F}^{2}\left(m_{B}^{2}-m_{D}^{2}\right)^{2}|\vec{p}_c|}{16 \pi m_{B}^{2}}\left|V_{u q}^{*} V_{c b}\right|^{2}\left|a_{1}\right|^{2} f_{P}^{2} F_{0}^{2}\left(m_{P}^{2}\right), \\
\Gamma\left(B\rightarrow D^{*} P\right) & =&\frac{G_{F}^{2}|\vec{p}_c|^{3}}{4 \pi}\left|V_{u q}^{*} V_{c b}\right|^{2}\left|a_{1}\right|^{2} f_{P}^{2} A_{0}^{2}\left(m_{P}^{2}\right), \\
\Gamma\left(B \rightarrow D V\right) & =&\frac{G_{F}^{2}|\vec{p}_c|^{3}}{4 \pi}\left|V_{u q}^{*} V_{c b}\right|^{2}\left|a_{1}\right|^{2} f_{V}^{2} F_{+}^{2}\left(m_{V}^{2}\right), \\
\Gamma\left(B \rightarrow D^{*} V\right) & =&\frac{G_{F}^{2}|\vec{p}_c|}{16 \pi m_{B}^{2}}\left|V_{u q}^{*} V_{c b}\right|^{2}\left(\left|H_{0}\right|^{2}+\left|H_{+}\right|^{2}+\left|H_{-}\right|^{2}\right), 
\en
where $\vec{p}_c$ is the momentum of either of the two final state mesons in the $B$ rest frame and $H_{0,\pm}$ are the helicity amplitudes,
\begin{align}
H_{0} &= \frac{\mathrm{i} f_{V}a_1}{2 m_{D^{*}}}\left[
    \left(m_{B}^{2} - m_{D^{*}}^{2} - m_{V}^{2}\right)\left(m_{B} + m_{D^{*}}\right) A_{1}^{B D^{*}}\left(m_{V}^{2}\right) \right. \nonumber  \\ 
    &\quad - \frac{4 m_{B}^{2} p_{c}^{2}}{m_{B} + m_{D^{*}}} A_{2}^{B D^{*}}\left(m_{V}^{2}\right)
    \left. \right], \\ 
H_{\pm} &= \mathrm{i} f_{V} m_{V}a_1\left[-
    \left(m_{B} + m_{D^{*}}\right) A_{1}^{BD^{*}}\left(m_{V}^{2}\right) \mp \frac{2 m_{B} p_{c}}{m_{B} + m_{D^{*}}} V^{BD^{\star}}\left(m_{V}^{2}\right)
    \right]. 
\end{align}
The polarization fractions are defined as 
\be
f_{L, \|, \perp}=\frac{H_{0, \|, \perp}}{H_{0}+H_{\|}+H_{\perp}}
\en
where $H_{\parallel}$ and $H_{\perp}$ are parallel and perpendicular amplitudes, respectively, and can be obtained through $H_{\parallel,\perp}=\frac{(H_{-}\pm H_{+})}{\sqrt{2}}$.

\section{Numerical results and discussions} \label{numer}
\subsection{Transition Form Factors}
\begin{table}[H]
\caption{The values of the input parameters\cite{DELPHI,LHCb2016,PDG2022,034035,862006,PDGT2024}. }
\label{tab:constant}
\begin{tabular*}{16.5cm}{@{\extracolsep{\fill}}l|cccccc}
  \hline\hline
\textbf{Mass(\text{GeV})} &$m_{b}=4.8$
&$m_{c}=1.4$&$m_{s}=0.37$&$m_{u,d}=0.25$   \\[1ex]
&$m_{\pi}=0.140$&$m_{K}=0.494$&$m_{\rho}=0.775$&$m_{K^{*}}=0.892$\\[1ex]
&$m_{D}=1.8695$& $m_{D(2S)}=2.549$& $m_{D_{s}}=1.96835$& $m_{D_{s}(2S)}=2.591$  \\[1ex]
& $m_{D_{s}^{*\pm}}=2.1066 $  &$m_{D^{*0}}=2.0068 $&$m_{D^{*\pm}}=2.0102 $& $ m_{B}=5.279$  \\[1ex]
& $ m_{D_{s}^{*\pm}(2S)}=2.732 $& $m_{D^{*0}(2S)}=2.627$&$m_{D^{*\pm}(2S)}=2.637$&$m_{\bar{B}_{s}^{0}}=5.367$\\[1ex]
\hline
\end{tabular*}
\begin{tabular*}{16.5cm}{@{\extracolsep{\fill}}l|ccccc}
  \hline
\multirow{2}{*}\textbf{ CKM}&$V_{cb}=0.0408\pm0.0014$&$V_{us}=0.2243\pm0.0008$&$V_{ud}=0.97373\pm0.00031$\\[1ex]
\hline
\end{tabular*}
\begin{tabular*}{16.5cm}{@{\extracolsep{\fill}}l|ccccc}
\hline
\textbf{ decay constants(\text{GeV})} & $f_{\pi}=0.132$ & $f_{K}=0.16$& $f_{\rho}=0.209$  \\[1ex]
&  $f_{K^{*}}=0.217$ &$f_{D_{s}}=0.2499\pm0.0005$ & $f_{D}=0.2058\pm0.0089$ \\[1ex]
& $f_{D_{s}(2S)}=0.161\pm0.020$ & $f_{D(2S)}=0.117\pm0.020$& $f_{D_{s}^{*\pm}}=0.272^{+0.039}_{-0.038}$   \\[1ex]
& $f_{D^{*0}}=0.339\pm0.022$& $f_{D^{\star\pm}}=0.341\pm0.023$ & $f_{D_{s}^{*\pm}(2S)}=0.312\pm0.017$ \\[1ex]
 & $f_{D^{*0}(2S)}=0.289\pm0.016$& $f_{D^{*\pm}(2S)}=0.290\pm0.016$ &$f_B=0.190\pm0.025$ \\[1ex]
&$f_{B_s}=0.253\pm0.008\pm0.007$&\\[1ex]
\hline\hline
\end{tabular*}
\begin{tabular*}{16.5cm}{@{\extracolsep{\fill}}l|ccccc}
\textbf{shape parameters(\text{GeV})}&$\beta^{\prime}_{D_{s}^{*\pm}(1S)}=0.436^{+0.039}_{-0.040}$&$\beta^{\prime}_{D^{\star0}(1S)}=0.500^{+0.140}_{-0.187}$&$\beta^{\prime}_{D}=0.466^{+0.022}_{-0.021}$\\[1ex]
& $\beta_{D_{s}^{*\pm}(2S)}^{'}=0.473^{+0.041}_{-0.041}$&$\beta^{\prime}_{D^{*0}(2S)}=0.456^{+0.004}_{-0.003}$&$\beta^{\prime}_{D_{s}}=0.5416^{+0.001}_{-0.002}$\\[1ex]
& $\beta_{D^{*\pm}(1S)}^{\prime}=0.502^{+0.041}_{-0.041}$&$\beta^{\prime}_{D^{*\pm}(2S)}=0.453^{+0.004}_{-0.003}$&$\beta^{\prime}_{\bar{B}_{s}^{0}}=0.626^{+0.045}_{-0.045}$\\[1ex]
& $\beta_{D(2S)}^{\prime}=0.297^{+0.041}_{-0.041}$&$\beta^{\prime}_{D_{s}(2S)}=0.422^{+0.026}_{-0.025}$&$\beta^{\prime}_{B}=0.555^{+0.060}_{-0.060}$\\[1ex]
\hline\hline
\end{tabular*}
\begin{tabular*}{16.5cm}{@{\extracolsep{\fill}}l|ccc}
\textbf{mean life$(10^{-12}s)$}&$\tau_{B^{0}}=1.517\pm0.004$&$\tau_{B^{0}_{s}}=1.516\pm0.006$&$\tau_{B^{\pm}}=1.638\pm0.004$ \\[1ex]
\hline\hline
\end{tabular*}
\end{table}
The input parameters, including the masses of the initial and final mesons (quarks), the CKM matrix elements, the lifetimes of the decaying mesons, 
the decay constants of related mesons and the shape parameters fitted by the decay constants are listed in Table \ref{tab:constant}. It is noticed that in the absence of the observed date on the masses and the decay constants of these first radially excited charmed and strange-charmed mesons, we take the corresponding predictions from established theoretical approaches \cite{034035,862006,PB2005,DB1311}.
Based on the input parameters listed in Table \ref{tab:constant}, the numerical results of the transition form factors at $q^2=0$ can be obtained, as shown in Tables \ref{table2} and \ref{table3}.

\begin{table}[H]
\caption{The $B_{(s)}\rightarrow D_{(s)}(1S,2S)$ transition form factors in the CLFQM.
The uncertainties are from the decay constants of $B_{(s)}$ and final state mesons.}
\begin{center}
\scalebox{1}{
\begin{tabular}{||c|c|c|c|c||}
\hline\hline
F&F(0)&$F(q^{2}_{max})$&$a$&$b$\\
\hline
$F^{BD}_{1}$&$0.66^{+0.00+0.01}_{-0.01-0.01}$&$0.81^{+0.01+0.01}_{-0.00-0.01}$&$0.80^{+0.01+0.04}_{-0.02-0.04}$&$0.86^{+0.01+0.02}_{-0.01-0.02}$\\
$F^{BD}_{0}$&$0.66^{+0.00+0.01}_{-0.01-0.01}$&$0.70^{+0.02+0.01}_{-0.03-0.01}$&$0.46^{+0.04+0.00}_{-0.02-0.01}$&$0.77^{+0.01+0.05}_{-0.01-0.05}$\\
\hline
$F^{BD(2S)}_{1}$&$0.26^{+0.01+0.01}_{-0.02-0.02}$&$0.34^{+0.02+0.01}_{-0.03-0.03}$&$0.99^{+0.04+0.16}_{-0.10-0.18}$&$0.66^{+0.07+0.17}_{-0.12-0.15}$\\
$F^{BD(2S)}_{0}$&$0.26^{+0.01+0.02}_{-0.01-0.02}$&$0.32^{+0.03+0.02}_{-0.00-0.04}$&$0.65^{+0.03+0.05}_{-0.01-0.04}$&$-0.24^{+0.02+0.01}_{-0.03-0.03}$\\
\hline\hline
$F^{B_{s}D_{s}}_{1}$&$0.67^{+0.00+0.01}_{-0.00-0.01}$&$0.81^{+0.00+0.00}_{-0.01-0.01}$&$0.82^{+0.00+0.02}_{-0.01-0.02}$&$0.96^{+0.01+0.03}_{-0.02-0.03}$\\
$F^{B_{s}D_{s}}_{0}$&$0.67^{+0.00+0.01}_{-0.00-0.01}$&$0.71^{+0.01+0.01}_{-0.02-0.01}$&$0.48^{+0.00+0.08}_{-0.01-0.08}$&$0.85^{+0.02+0.06}_{-0.02-0.06}$\\
\hline
$F^{B_{s}D_{s}(2S)}_{1}$&$0.26^{+0.02+0.02}_{-0.02-0.02}$&$0.30^{+0.02+0.01}_{-0.02-0.02}$&$0.58^{+0.12+0.05}_{-0.16-0.01}$&$0.33^{+0.11+0.04}_{-0.13-0.02}$\\
$F^{B_{s}D_{s}(2S)}_{0}$&$0.26^{+0.01+0.01}_{-0.02-0.02}$&$0.25^{+0.02+0.01}_{-0.02-0.02}$&$-0.10^{+0.22+0.22}_{-0.14-0.26}$&$-0.07^{+0.13+0.18}_{-0.21-0.08}$\\
\hline\hline
\end{tabular}}\label{table2}
\end{center}
\end{table}

All computations are carried out within the reference frame $q^+=0$, where the form factors can only be obtained at spacelike momentum transfers $q^2=-q^2_{\bot}\leq0$. We need to know the form factors in the timelike region for the physical decay processes. Here we use the
following double-pole approximation to parametrize the form factors in the spacelike region and then extend to the timelike region,
\be
F\left(q^{2}\right)=\frac{F(0)}{1-a q^{2} / m^{2}+b q^{4} / m^{4}},
\en
where $m$ represents the initial meson mass and $F(q^{2})$ denotes the different form factors $F_{1},F_{0},V,A_{0},A_{1}$ and $A_{2}$.
The values of $a$ and $b$ can be obtained by performing a 3-parameter fit to the form factors in the range $-15 \text{GeV}^2\leq q^2\leq0$, which are collected in Tables \ref{table2} and \ref{table3}. The uncertainties arise from the decay constants of the initial and the final mesons.

\begin{table}[H]
	\caption{The $B\rightarrow D^{*}(1S,2S)$ and $B_{s}\rightarrow D^{*}_{s}(1S,2S)$ transition form factors in the CLFQM.
		The uncertainties are from the decay constants of $B_{(s)}$ and final state mesons.}
	\begin{center}
		\scalebox{1}{
			\begin{tabular}{||c|c|c|c|c||}
				\hline\hline
				F&F(0)&$F(q^{2}_{max})$&$a$&$b$\\
				\hline
				$V^{BD^{*}}$&$0.77^{+0.00+0.01}_{-0.00-0.01}$&$0.94^{+0.00+0.01}_{-0.00-0.01}$&$0.78^{+0.00+0.03}_{-0.01-0.03}$&$0.82^{+0.04+0.14}_{-0.08-0.16}$\\
				$A_{0}^{BD^{*}}$&$0.75^{+0.00+0.06}_{-0.01-0.02}$&$0.79^{+0.00+0.05}_{-0.02-0.02}$&$0.17^{+0.01+0.01}_{-0.01-0.00}$&$0.12^{+0.07+0.02}_{-0.06-0.03}$\\
				$A_{1}^{BD^{*}}$&$0.67^{+0.00+0.01}_{-0.01-0.11}$&$0.76^{+0.00+0.01}_{-0.01-0.01}$&$0.38^{+0.01+0.01}_{-0.01-0.01}$&$0.19^{+0.05+0.09}_{-0.05-0.09}$\\
				$A_{2}^{BD^{*}}$&$0.58^{+0.00+0.00}_{-0.01-0.02}$&$0.69^{+0.01+0.01}_{-0.01-0.02}$&$0.68^{+0.02+0.00}_{-0.02-0.00}$&$0.66^{+0.25+0.10}_{-0.12-0.16}$\\
				\hline
			$V^{BD^{*}(2S)}$&$0.19^{+0.05+0.01}_{-0.06-0.01}$&$0.19^{+0.01+0.03}_{-0.04-0.01}$&$0.11^{+0.05+0.03}_{-0.06-0.05}$&$0.34^{+0.00+0.03}_{-0.04-0.06}$\\
			$A_{0}^{BD^{*}(2S)}$&$0.27^{+0.04+0.01}_{-0.05-0.00}$&$0.28^{+0.00+0.01}_{-0.02-0.00}$&$0.24^{+0.05+0.00}_{-0.06-0.00}$&$-0.07^{+0.15+0.03}_{-0.29-0.06}$\\
			$A_{1}^{BD^{*}(2S)}$&$0.16^{+0.04+0.01}_{-0.05-0.00}$&$0.15^{+0.05+0.02}_{-0.04-0.02}$&$-0.23^{+0.04+0.00}_{-0.04-0.00}$&$0.17^{+0.03+0.01}_{-0.01-0.01}$\\
			$A_{2}^{BD^{*}(2S)}$&$-0.05^{+0.04+0.01}_{-0.04-0.00}$&$0.01^{+0.00+0.00}_{-0.00-0.00}$&$-1.10^{+0.02+0.01}_{-0.01-0.01}$&$-0.63^{+0.02+0.01}_{-0.01-0.00}$\\
				\hline\hline
					$V^{B_{s}D_{s}^{*}}$&$0.78^{+0.01+0.01}_{-0.01-0.01}$&$0.87^{+0.00+0.00}_{-0.00-0.01}$&$0.86^{+0.01+0.04}_{-0.01-0.04}$&$1.11^{+0.01+0.02}_{-0.01-0.02}$\\
				$A_{0}^{B_{s}D_{s}^{*}}$&$0.74^{+0.01+0.01}_{-0.01-0.01}$&$0.69^{+0.01+0.01}_{-0.01-0.00}$&$0.23^{+0.01+0.01}_{-0.01-0.01}$&$0.21^{+0.00+0.01}_{-0.00-0.01}$\\
				$A_{1}^{B_{s}D_{s}^{*}}$&$0.66^{+0.01+0.01}_{-0.02-0.02}$&$0.95^{+0.00+0.00}_{-0.00-0.01}$&$0.81^{+0.01+0.02}_{-0.01-0.02}$&$0.93^{+0.01+0.01}_{-0.01-0.01}$\\
				$A_{2}^{B_{s}D_{s}^{*}}$&$0.57^{+0.00+0.00}_{-0.01-0.00}$&$0.68^{+0.01+0.00}_{-0.01-0.00}$&$0.80^{+0.00+0.03}_{-0.01-0.03}$&$0.96^{+0.01+0.03}_{-0.01-0.02}$\\
				\hline
				$V^{B_{s}D_{s}^{*}(2S)}$&$0.26^{+0.03+0.04}_{-0.03-0.04}$&$0.28^{+0.00+0.02}_{-0.09-0.10}$&$0.25^{+0.02+0.04}_{-0.01-0.04}$&$0.30^{+0.03+0.04}_{-0.01-0.00}$\\
				$A_{0}^{B_{s}D_{s}^{*}(2S)}$&$0.31^{+0.02+0.01}_{-0.03-0.02}$&$0.33^{+0.00+0.08}_{-0.01-0.07}$&$0.21^{+0.01+0.03}_{-0.05-0.00}$&$-0.09^{+0.04+0.10}_{-0.02-0.14}$\\
				$A_{1}^{B_{s}D_{s}^{*}(2S)}$&$0.21^{+0.02+0.03}_{-0.03-0.03}$&$0.20^{+0.00+0.01}_{-0.07-0.06}$&$-0.16^{+0.20+0.04}_{-0.17-0.02}$&$0.12^{+0.00+0.12}_{-0.01-0.07}$\\
				$A_{2}^{B_{s}D_{s}^{*}(2S)}$&$-0.01^{+0.02+0.06}_{-0.02-0.06}$&$0.01^{+0.00+0.03}_{-0.01-0.01}$&$-4.03^{+0.51+0.75}_{-0.47-1.38}$&$4.31^{+0.00+2.13}_{-0.00-2.44}$\\
				\hline\hline
		\end{tabular}}\label{table3}
	\end{center}
\end{table}

An estimate of the $SU(3)_F$ breaking effects can be obtained by studying the form factor ratios between the transitions $B_{s} \to D^{(*)}_{s}(1S,2S)$ and $B \to D^{(*)}(1S,2S)$, and the results are listed as 
\be
\frac{F_{1}(B_{s} \to D_{s})}{F_{1}(B \to D)}&=&1.02,
\frac{F_{1}(B_{s} \to D_{s}(2S))}{F_{1}(B \to D(2S))}=0.76,  \\
\frac{V(B_{s} \to D^{*}_{s})}{V(B \to D^{*})}&=&1.01,
\frac{V(B_{s} \to D^{*}_{s}(2S))}{V(B \to D^{*}(2S))}=1.37,  \\
\frac{A_{1}(B_{s} \to D^{*}_{s})}{A_{1}(B \to D^{*})}&=&0.99,
\frac{A_{1}(B_{s} \to D^{*}_{s}(2S))}{A_{1}(B \to D^{*}(2S))}=1.31,  \\
\frac{A_{2}(B_{s} \to D^{*}_{s})}{A_{2}(B \to D^{*})}&=&0.98,
\frac{A_{2}(B_{s} \to D^{*}_{s}(2S))}{A_{2}(B \to D^{*}(2S))}=0.20. 
\en

In Tables \ref{tableC} and \ref{form}, we compare the form factors of the transitions $B_{(s)}\to D_{(s)}, D^{*}_{(s)}$ at maximum recoil ($q^{2}=0$) with those obtained within the QCD LCSRs \cite{Wu:2025kdc,Fu:2013wqa,Zhang:2017rwz,Gao:2021sav,Zhong:2018exo,Li:2009wq}, the LQCD \cite{Yao:2019vty,MILC:2015uhg,Na:2015kha}, the RQM \cite{Faustov:2022ybm,Faustov:2012mt}, the QCDSR \cite{Blasi:1993fi}, the PQCD  \cite{Hu:2019bdf,Fan:2013qz, Fan:2013kqa}, the CCQM \cite{Soni:2021fky} and the previous CLFQM \cite{Verma:2011yw,GLI}. Upon comparison, we find that our predictions for the form factors of the transitions $B_{(s)}\to D_{(s)}, D^{*}_{(s)}$ are consistent with most of the other theoretical results, especially for the values of $F^{BD}_{1}$. In a word, these dramatically different theoretical calculations lead to rather close values of the transition form factors. Unfortunately, research on the $B_{(s)}\to D_{(s)}(2S), D^{*}_{(s)}(2S)$ transition form factors is still very limited, and we look forward to more studies on them in the future.
\begin{table}[H]
	\caption{The $B\to D$ transition form factor. As a comparison, we also present other theoretical results.}
	\begin{center}
		\scalebox{1}{
			\begin{tabular}{ccccccc}
				\hline
				&&$$&$$&$$&$$&$F^{BD}_{1}$\\
				\hline
				This work &&$$&$$&$$&$$&$0.66^{+0.01}_{-0.01}$\\
			LCSR \cite{Wu:2025kdc}&&$$&$$&$$&$$&$0.648^{+0.067}_{-0.063}$\\
				MC \cite{Wu:2025kdc}$^{a}$&&$$&$$&$$&$$&$0.652\pm0.023$\\
				PQCD \cite{Fan:2013qz}&&$$&$$&$$&$$&$0.52^{+0.12}_{-0.10}$\\
				LCSR \cite{Fu:2013wqa}&&$$&$$&$$&$$&$0.653^{+0.004}_{-0.011}$\\
				LCSR \cite{Zhang:2017rwz}&&$$&$$&$$&$$&$0.673^{+0.038}_{-0.041}$\\
				LCSR \cite{Gao:2021sav}&&$$&$$&$$&$$&$0.552\pm0.216$\\
				LCSR \cite{Zhong:2018exo}&&$$&$$&$$&$$&$0.659^{+0.029}_{-0.032}$\\
				LQCD \cite{Yao:2019vty}&&$$&$$&$$&$$&$0.658\pm0.017$\\
				LQCD \cite{MILC:2015uhg}&&$$&$$&$$&$$&$0.672\pm0.027$\\
				HPQCD \cite{Na:2015kha}&&$$&$$&$$&$$&$0.664\pm0.034$\\
				RQM \cite{Faustov:2022ybm}&&$$&$$&$$&$$&$0.696$\\
				\hline
		\end{tabular}}\label{tableC}
	\end{center}
    {\footnotesize $^a$ refers to the Mote Carlo method.}
\end{table}

\begin{figure}[H]
\vspace{0.30cm}
  \centering
  \subfigure[]{\includegraphics[width=0.38\textwidth]{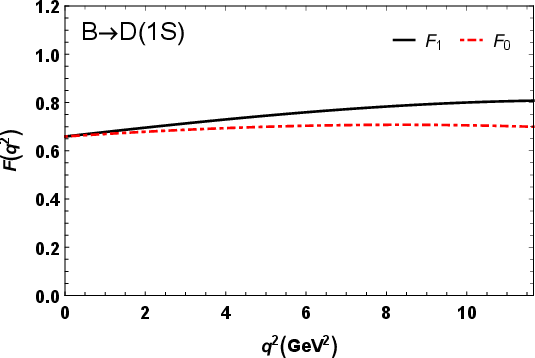}\quad}
  \subfigure[]{\includegraphics[width=0.38\textwidth]{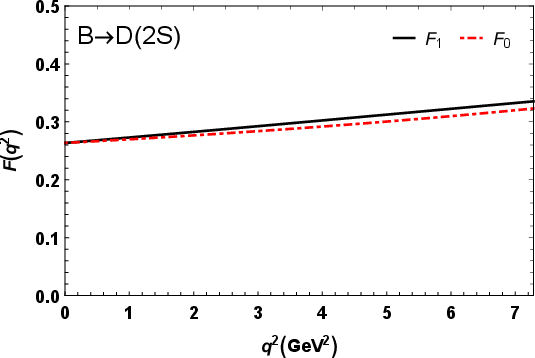}}\\
  \subfigure[]{\includegraphics[width=0.38\textwidth]{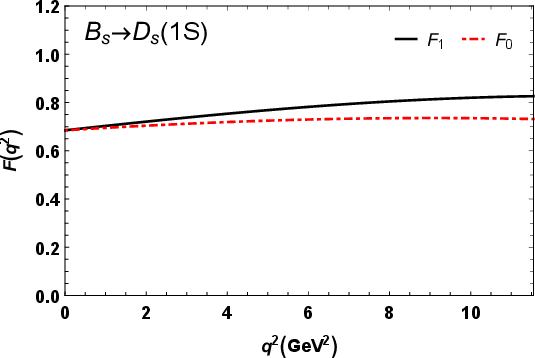}\quad}
  \subfigure[]{\includegraphics[width=0.38\textwidth]{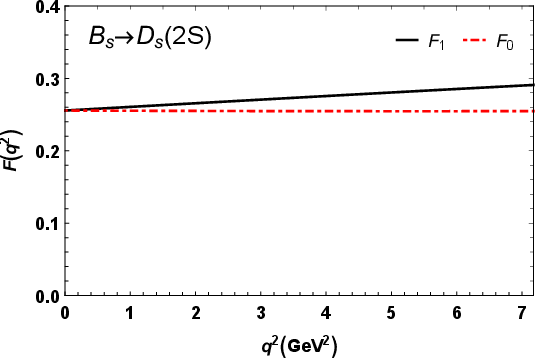}}\\
\caption{Form factors $F_{1}(q^2)$ and $F_{0}(q^2)$ of the transitions $B_{(s)}\rightarrow D_{(s)}(1S,2S)$.}\label{fig:T55}
\end{figure}

\begin{figure}[H]
	\vspace{0.50cm}
	\centering
	\subfigure[]{\includegraphics[width=0.38\textwidth]{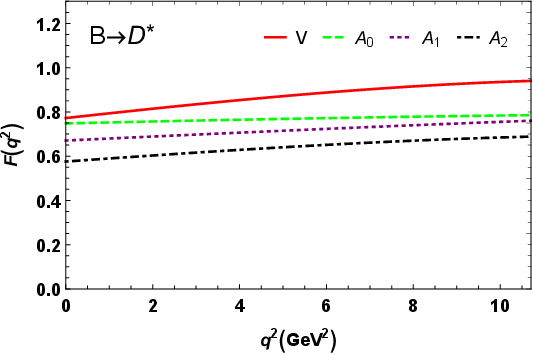}\quad}
	\subfigure[]{\includegraphics[width=0.38\textwidth]{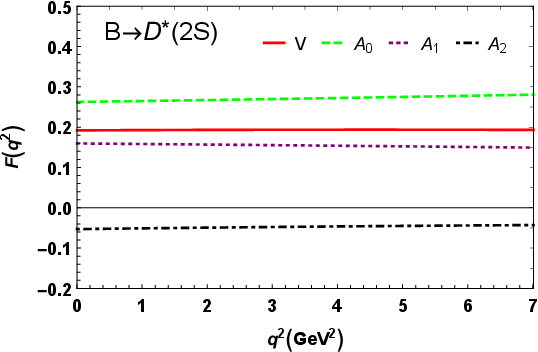}}\\
	\subfigure[]{\includegraphics[width=0.38\textwidth]{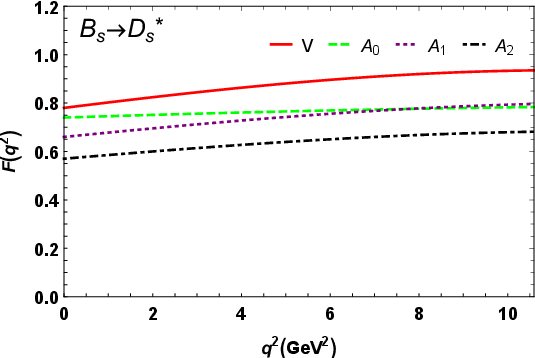}\quad}
	\subfigure[]{\includegraphics[width=0.38\textwidth]{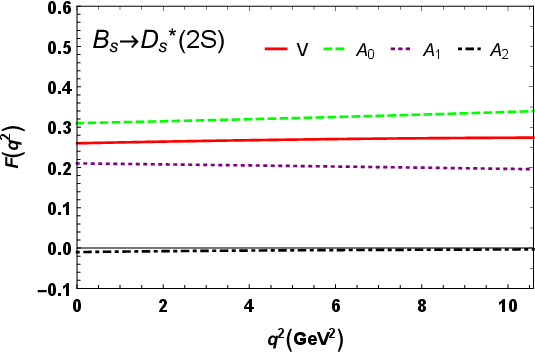}}
	\caption{Form factors $V(q^2)$, $A_{0}(q^2)$, $A_{1}(q^2)$ and $A_{2}(q^2)$ of the transitions $B_{(s)}\rightarrow D^{*}_{(s)}(1S,2S)$.}\label{fig:T5}
\end{figure}

We plot the $q^2$-dependence of the $B_{(s)}\to D_{(s)}(1S,2S)$ and $ B_{(s)}\to D^{*}_{(s)}(1S,2S)$ transition form factors in Figures \ref{fig:T55} and \ref{fig:T5}. In Figure \ref{fig:T55}, the form factors of the transitions $B_{(s)}\to D_{(s)}$ are approximately $2.5$ times larger than those of the transitions $B_{(s)}\to D_{(s)}(2S)$. The difference in the form factors between the transitions $B_{(s)}\to D^*_{(s)}$ and $B_{(s)}\to D^*_{(s)}(2S)$ is also very obvious, as shown in Figure \ref{fig:T5}.

\begin{table}[H]
	\caption{The $B\to D^*$ and $B_s\to D^{(*)}_s$ transition form factors at $q^{2}= 0$, together with other theoretical results for comparison.}
	\begin{center}
    \scalebox{0.8}{
		\begin{tabular}{|c|c|ccccc|c|}
			\hline\hline
			Transition  &Reference&\;\;$F_{0}(0)\;\;$&\;\;$V(0)$&\;\;$A_{0}(0)$&\;\;$A_{1}(0)$&\;\;$A_{2}(0)$\\
			\hline
			$B\rightarrow D^{*} $&This work&$-$&$0.77$&$0.75$&$0.67$&$0.58$\\
			\hline
			$ $&RQM \cite{Faustov:2022ybm}&$-$&$0.92$&$0.81$&$0.73$&$0.63$\\
				$ $&CLFQM \cite{Verma:2011yw}&$-$&$0.77$&$0.68$&$0.60$&$0.61$\\
			\hline
			$B_{s}\rightarrow D^{(*)}_{s} $&This work&$0.67$&$0.78$&$0.74$&$0.66$&$0.57$\\
			\hline
			$ $&RQM \cite{Faustov:2012mt}&$0.74$&$0.95$&$0.67$&$0.70$&$0.75$\\
			$ $&CLFQM \cite{Verma:2011yw}&$0.67$&$0.75$&$0.66$&$0.62$&$0.57$\\
			$ $&QCDSR \cite{Blasi:1993fi}&$0.70$&$0.63$&$0.52$&$0.62$&$0.75$\\
			$$&BS \cite{XJCHENHF}&$0.57$&$0.70$&$-$&$0.65$&$0.67$\\
			$ $&LCSR \cite{Li:2009wq}&$0.86$&$$&$-$&$-$&$-$\\
			$ $&LFQM \cite{GLI}&$-$&$0.74$&$0.63$&$0.61$&$0.59$\\
			$ $&PQCD \cite{Hu:2019bdf}&$0.52$&$0.64$&$0.48$&$0.50$&$0.53$\\
			$ $&PQCD \cite{Fan:2013kqa}&$0.55$&$0.62$&$0.47$&$0.49$&$0.52$\\
			$ $&CCQM \cite{Soni:2021fky}&$0.77$&$0.74$&$0.72$&$0.68$&$0.63$\\
			$ $&RQM \cite{Faustov:2022ybm}&$0.66$&$0.93$&$0.63$&$0.67$&$0.72$\\
			\hline\hline
		\end{tabular}}\label{form}
	\end{center}
\end{table}

\subsection{Nonleptonic decays}
In this section, using the form factors obtained in the preceding section, we give the branching ratios of the nonleptonic decays $B\to D(1S,2S)(\pi, \rho, K^{(*)}),$ $B_s\to D_s(1S,2S)(\pi, \rho, K^{(*)}),  B\to D^{*}(1S, 2S)(\pi, \rho, K^{(*)})$ and $B_{s}\to D^{*}_{s}(1S,2S)(\pi, \rho, K^{(*)})$, which are listed in Tables \ref{b1}, \ref{b2}, \ref{b3} and \ref{b4}, respectively. Here, uncertainties arise from the $B_{(s)}$ meson lifetime, the decay constants of initial and final state mesons, respectively. We compare our predictions with the results from the BS method \cite{Zhou:2020ijj}, the RQM \cite{Faustov:2012mt}, the QCDSRs \cite{Blasi:1993fi}, the RCQM \cite{Chen:2011ut}, the LCSRs \cite{Li:2009wq}, the three-point QCDSRs \cite{KRF}, the relativistic independent quark model (RIQM) \cite{Dash:2023hjr} and the PQCD approach \cite{lirh}. The available experimental data \cite{PDGT, JBEA} are also included. We adopt the Wilson coefficient $a_{1}=1.1$ in the calculations. The following are some comments.
\begin{enumerate}
\item
From Table \ref{b1}, one can find that the branching ratios of the decays $B\to D(1S)(\pi, \rho, K^{(*)})$ fall within the range of $10^{-4}\sim10^{-3}$, which are about one order larger than those of the corresponding decays $B\to D(2S)(\pi, \rho, K^{(*)})$. The small branching ratios for the latter are related to the node structure of the wave function of $D(2S)$ meson. When calculated the overlap integral of the wave functions, contributions from the positive and negative parts of the D(2S) meson wave function cancel each other out, which result in the small decay branching fractions. 
\begin{table}[H]
	\caption{The branching ratios of the decays $B\to D(1S,2S)(\pi,\rho,K^{(*)})$.}
	\begin{center}
		\scalebox{1}{
			\begin{tabular}{|c|c|c|c|c|c|c|c|c|}
				\hline\hline
				Modes&This work&\cite{Zhou:2020ijj}&\cite{lirh}&\cite{Huber:2016xod}&\cite{Chang:2017sdl}&Exp.\cite{PDGT2024}&Unit\\
				\hline
				$\bar{B}^{0}\to D^{+}(1S) \pi^{-}$&$4.36^{+0.02+0.01+0.12}_{-0.01-0.08-0.12}$&$3.24$&$2.69$&$3.93$&$3.582$&$2.51$&$$\\
				$\bar{B}^{0}\to D^{+}(1S) K^{-}$&$0.34^{+0.00+0.00+0.01}_{-0.00-0.01-0.01}$&$0.245$&$0.243$&$0.301$&$0.2712$&$0.205$&$(10^{-3})$\\
				$\bar{B}^{0}\to D^{+}(1S) \rho^{-}$&$10.09^{+0.04+0.02+0.27}_{-0.01-0.02-0.27}$&$7.91$&$6.96$&$10.42$&$-$&$7.6$&$$\\
				$\bar{B}^{0}\to D^{+}(1S) K^{*-}$&$0.56^{+0.00+0.00+0.01}_{-0.00-0.01-0.02}$&$0.431$&$0.407$&$0.525$&$-$&$0.45$&$$\\
				\hline
				$\bar{B}^{0}\to D^{+}(2S) \pi^{-}$&$4.76^{+0.01+0.48+0.33}_{-0.01-0.62-0.57}$&$0.458$&$-$&$-$&$-$&$-$&$$\\
				$\bar{B}^{0}\to D^{+}(2S) K^{-}$&$0.37^{+0.00+0.04+0.04}_{-0.00-0.05-0.03}$&$0.034$&$-$&$-$&$-$&$-$&$(10^{-4})$\\
				$\bar{B}^{0}\to D^{+}(2S) \rho^{-}$&$10.54^{+0.03+1.06+0.73}_{-0.03-1.37-1.26}$&$1.03$&$-$&$-$&$-$&$-$&$$\\
				$\bar{B}^{0}\to D^{+}(2S) K^{*-}$&$0.58^{+0.00+0.06+0.04}_{-0.00-0.08-0.07}$&$0.0557$&$-$&$-$&$-$&$-$&$$\\
				\hline\hline
				$B^{-}\to D^{0}(1S) \pi^{-}$&$4.71^{+0.01+0.00+0.12}_{-0.01-0.01-0.13}$&$3.49$&$5.11$&$-$&$-$&$4.61$&$$\\
				$B^{-}\to D^{0}(1S) K^{-}$&$0.37^{+0.00+0.00+0.01}_{-0.00-0.01-0.01}$&$0.264$&$0.40$&$-$&$-$&$0.364$&$(10^{-3})$\\
				$B^{-}\to D^{0}(1S) \rho^{-}$&$10.88^{+0.03+0.00+0.27}_{-0.03-0.02-0.31}$&$8.40$&$11.3$&$-$&$-$&$9.7$&$$\\
				$B^{-}\to D^{0}(1S) K^{*-}$&$0.60^{+0.00+0.00+0.02}_{-0.00-0.01-0.02}$&$0.466$&$0.649$&$-$&$-$&$0.53$&$$\\
				\hline
				$B^{-}\to D^{0}(2S) \pi^{-}$&$5.14^{+0.01+0.67+0.36}_{-0.01-0.67-0.61}$&$0.488$&$-$&$-$&$-$&$-$&$$\\
				$B^{-}\to D^{0}(2S) K^{-}$&$0.40^{+0.00+0.05+0.03}_{-0.00-0.05-0.05}$&$0.0362$&$-$&$-$&$-$&$-$&$(10^{-4})$\\
				$B^{-}\to D^{0}(2S) \rho^{-}$&$11.36^{+0.03+1.48+0.79}_{-0.03-1.48-1.35}$&$1.09$&$-$&$-$&$-$&$-$&$$\\
				$B^{-}\to D^{0}(2S) K^{*-}$&$0.62^{+0.00+0.08+0.04}_{-0.00-0.08-0.07}$&$0.0595$&$-$&$-$&$-$&$-$&$$\\
				\hline\hline
			\end{tabular}\label{b1}}
	\end{center}
\end{table}
\item
The branching ratios of the neutral decays $\bar B^0_{(s)}\to D_{(s)}(1S)^+(\pi, \rho, K^{(*)})^-$ are larger than the experimental data \cite{PDGT}, as shown in Tables \ref{b1} and \ref{b2}. A similar situation also occurs in the calculations from the QCDF approach in Refs. \cite{Huber:2016xod,Chang:2017sdl}, where although the contributions from more comprehensive amplitudes \cite{Chang:2017sdl} and the next-to-next-to-leading-order (NNLO) vertex corrections \cite{Huber:2016xod} were considered, the predictions for the 
branching raitos of the decays $\bar B_{(s)}^0\to D_{(s)}(1S)^+(\pi, \rho, K^{(*)})^-$ still exceed the data. By comparison, the results from the PQCD approach given in \cite{lirh} are closer to the experimental data for these neutral $B_{(s)}$ decays, where except for the factorizable emission diagram amplitude, the nonfactorizable emission diagram amplitude may also give a significant contribution. Furthermore, these two kinds of amplitudes partially cancel each other.  It is strange that the difference between our predictions and expeimental data for the charged decays $B^-\to D(1S)^0(\pi,\rho, K^{(*)})^-$ is not significant.
\begin{table}[H]
	\caption{The branching ratios of the decays $\bar{B}^{0}_{s}\to D^{+}_{s}(1S,2S)(\pi, \rho, K^{(*)})^-$, together with other theoretical results and data for comparison.}
	\begin{center}
		\scalebox{0.8}{
			\begin{tabular}{||c|c|c|c|c|c|c|c|c|c|c|c|c|c||}
				\hline\hline
				Modes&This work&\cite{Zhou:2020ijj}&\cite{Faustov:2012mt}&\cite{Albertus:2014eqa}&\cite{Blasi:1993fi}&\cite{Chen:2011ut}&\cite{lirh}&\cite{KRF}&\cite{Li:2009wq}&\cite{Chang:2017sdl}&\cite{Huber:2016xod}&Exp.\cite{PDGT2024}&Unit\\
				\hline
				$\bar{B}_{s}^{0}\to D_{s}^{+}(1S) \pi^{-}$&$4.56^{+0.01+0.45+0.17}_{-0.01-0.31-0.15}$&$2.92$&$3.5$&$5.3$&$5$&$2.7$&$2.13$&$1.42$&$1.7$&$4.717$&$4.39$&$2.98\pm0.14$&$$\\
				$\bar{B}_{s}^{0}\to D_{s}^{+}(1S) K^{-}$&$0.36^{+0.00+0.04+0.01}_{-0.00-0.02-0.01}$&$0.221$&$0.28$&$0.4$&$0.4$&$0.21$&$0.171$&$0.103$&$0.13$&$0.3575$&$0.334$&$0.225\pm0.012$&$(10^{-3})$\\
				$\bar{B}_{s}^{0}\to D_{s}^{+}(1S) \rho^{-}$&$10.52^{+0.03+1.06+0.42}_{-0.03-0.76-0.38}$&$7.04$&$9.4$&$12.6$&$13$&$6.4$&$5.1$&$-$&$4.2$&$-$&$11.30$&$6.8\pm1.4$&$$\\
				$\bar{B}_{s}^{0}\to D_{s}^{+}(1S) K^{*-}$&$0.58^{+0.00+0.06+0.02}_{-0.00-0.04-0.02}$&$0.392$&$0.47$&$0.8$&$0.6$&$0.38$&$0.302$&$0.05$&$0.24$&$-$&$0.564$&$-$&$$\\
				\hline
				$\bar{B}_{s}^{0}\to D_{s}^{+}(2S) \pi^{-}$&$4.69^{+0.02+0.82+0.67}_{-0.02-0.47-0.43}$&$1.13$&$7$&$-$&$-$&$-$&$-$&$-$&$-$&$-$&$-$&$-$&$$\\
				$\bar{B}_{s}^{0}\to D_{s}^{+}(2S) K^{-}$&$0.36^{+0.00+0.06+0.04}_{-0.00-0.05-0.03}$&$0.084$&$0.5$&$-$&$-$&$-$&$-$&$-$&$-$&$-$&$-$&$-$&$(10^{-4})$\\
				$\bar{B}_{s}^{0}\to D_{s}^{+}(2S) \rho^{-}$&$10.41^{+0.04+1.87+1.53}_{-0.04-1.00-0.09}$&$2.49$&$17$&$-$&$-$&$-$&$-$&$-$&$-$&$-$&$-$&$-$&$$\\
				$\bar{B}_{s}^{0}\to D_{s}^{+}(2S) K^{*-}$&$0.57^{+0.00+0.10+0.08}_{-0.00-0.05-0.05}$&$0.134$&$0.8$&$-$&$-$&$-$&$-$&$-$&$-$&$-$&$-$&$-$&$$\\
				\hline\hline
			\end{tabular}\label{b2}}
	\end{center}
\end{table}
\item 
Our predictions for the branching ratios of the decays with ground state $D_{(s)}$ or $D^*_{(s)}$ meson involved are comparable with those given by the BS method \cite{Zhou:2020ijj}. While if replaced the ground state charmed meson with the radially excited one in these decays, the predicted results between these two approaches show significant discrepancies. For example, $Br(B\to D(2S)(\pi, \rho, K^{(*)}))$ are about one order larger than those given by the BS method \cite{Zhou:2020ijj}. Such significant difference can be clarified by future experiments. 
\item A similar situation also occurs in the decays $B\to D^*(1S,2S)(\pi, \rho, K^{(*)})$ as shown in Table \ref{b3}, where the branching ratios of the neutral decays $\bar B^0\to D^{*0}(1S)(\pi, \rho, K^{(*)})$ are about 2 times as large as the data, while the difference between our predictions and experimental measurements for the charged decays $B^-\to D^{*0}(1S)(\pi, \rho, K^{(*)})^-$ is not significant. Our predictions are comparable with those given by the BS approach \cite{Zhou:2020ijj} for the ground state $D^*(1S)$ case, but much larger for the excited state $D^*(2S)$ case. Note that our results are consistent well with the RIQM calculations \cite{Dash:2023hjr} in both $D^*(1S)$ and $D^*(2S)$ cases. 
\item Compared to the experimental data, the similar situation occurs again in the decays $\bar B^0_s\to D^{*+}_s(1S)(\pi, \rho, K)^-$, that is the predictions for their branching ratios have an obvious exceedance, as shown in Table \ref{b4}. Certainly, our predictions for the branching ratios of the decays$\bar B^0_s\to D^{*+}_s(2S)(\pi, \rho, K^{(*)})^-$ are also larger than those given by the BS approach \cite{Zhou:2020ijj}, but agree well the RQM \cite{Faustov:2012mt} and the RIQM results \cite{Dash:2023hjr}.
 \item 
 When the same initial state $B_{(s)}$ decays to the same final states $D^{(*)}_{(s)}(nS), n=1,2$ with emission of different types of light mesons, the branching ratios exhibit a clear hierarchical pattern, primarily due to the hierarchical structure of the CKM factors, that is $V_{ud}\gg V_{us}$,
\begin{footnotesize}
\begin{equation}
\begin{aligned}
\mathcal{B} r\left(B \rightarrow D(nS) \pi\right) &\gg \mathcal{B} r\left(B\rightarrow D(nS) K\right) , \mathcal{B} r\left(B \rightarrow D(nS) \rho\right) \gg \mathcal{B} r\left(B \rightarrow D(nS) K^{*}\right), \\
\mathcal{B} r\left(B \rightarrow D^{*}(nS) \pi\right) &\gg \mathcal{B} r\left(B \rightarrow D^{*}(nS) K\right) , \mathcal{B} r\left(B\rightarrow D^{*}(nS) \rho\right) \gg \mathcal{B} r\left(B \rightarrow D^{*}(nS) K^{*}\right) .\non
\end{aligned}
\end{equation}
\end{footnotesize}
The upper relationships remain valid when $B$ and $D^{(*)}$ are replaced with $B_s$ and $D^{(*)}_s$, respectively.
\end{enumerate}
\begin{table}[H]
	\caption{The branching ratios $(10^{-3})$ of the decays $B\to D^*(1S,2S)(\pi, \rho, K^{(*)})$, together with other theoretical results and data for comparison.}
	\begin{center}
		\scalebox{1}{
			\begin{tabular}{|c|c|c|c|c|}
				\hline\hline
			Modes&$\bar{B}^{0}\to D^{*+}(1S) \pi^{-}$&$\bar{B}^{0}\to D^{*+}(1S) K^{-}$&$\bar{B}^{0}\to D^{*+}(1S) \rho^{-}$&$\bar{B}^{0}\to D^{*+}(1S) K^{*-}$\\
			\hline
				This work&$5.25^{+0.02+0.06+0.18}_{-0.01-0.02-0.42}$&$0.40^{+0.00+0.00+0.00}_{-0.00-0.01-0.03}$&$14.32^{+0.06+0.15+0.05}_{-0.02-0.48-0.07}$&$0.83^{+0.00+0.01+0.00}_{-0.00-0.03-0.06}$\\
				\cite{Zhou:2020ijj}	&$3.80$&$0.281$&$8.73$&$0.758$\\
				\cite{Dash:2023hjr}&$-$&$-$&$14.24$&$0.83$\\
                \cite{lirh}& $2.60$ & $0.237$ & $7.94$ & $0.488$\\
				Exp.\cite{PDGT2024}	&$2.66$&$0.216$&$6.8$&$0.33$\\
				\hline
				Modes&$\bar{B}^{0}\to D^{*+}(2S) \pi^{-}$&$\bar{B}^{0}\to D^{*+}(2S) K^{-}$&$\bar{B}^{0}\to D^{*+}(2S) \rho^{-}$&$\bar{B}^{0}\to D^{*+}(2S) K^{*-}$\\
				\hline
				This work&$0.45^{+0.00+0.16+0.01}_{-0.00-0.16-0.01}$&$0.03^{+0.00+0.01+0.00}_{-0.00-0.01-0.00}$&$1.09^{+0.00+0.41+0.02}_{-0.00-0.41-0.02}$&$0.06^{+0.00+0.02+0.00}_{-0.00-0.02-0.00}$\\
					\cite{Zhou:2020ijj}	&$0.038$&$0.00274$&$0.0267$&$0.00162$\\
					\cite{Dash:2023hjr}&$-$&$-$&$1.12$&$0.07$\\
				\hline\hline
				Modes&$B^{-}\to D^{*0}(1S) \pi^{-}$&$B^{-}\to D^{*0}(1S) K^{-}$&$B^{-}\to D^{*0}(1S) \rho^{-}$&$B^{-}\to D^{*0}(1S) K^{*-}$\\
				\hline
				This work&$5.67^{+0.01+0.05+0.98}_{-0.01-0.19-2.53}$&$0.43^{+0.00+0.00+0.07}_{-0.00-0.01-0.19}$&$15.45^{+0.04+0.13+2.30}_{-0.04-0.53-6.52}$&$0.89^{+0.00+0.01+0.13}_{-0.00-0.03-0.37}$\\
				\cite{Zhou:2020ijj}	&$4.11$&$0.304$&$8.73$&$0.846$\\
					\cite{Dash:2023hjr}&$-$&$-$&$15.39$&$0.90$\\
                    \cite{lirh} & $5.04$& $0.398$ &$11.7$ & $0.682$ \\
			Exp.\cite{PDGT2024}	&$5.17$&$0.419$&$9.8$&$0.81$\\
				\hline
				Modes	&$B^{-}\to D^{*0}(2S) \pi^{-}$&$B^{-}\to D^{*0}(2S) K^{-}$&$B^{-}\to D^{*0}(2S) \rho^{-}$&$B^{-}\to D^{*0}(2S) K^{*-}$\\
				\hline
				This work&$0.46^{+0.00+0.16+0.01}_{-0.00-0.17-0.01}$&$0.03^{+0.00+0.01+0.00}_{-0.00-0.01-0.00}$&$1.10^{+0.00+0.42+0.02}_{-0.00-0.42-0.02}$&$0.06^{+0.00+0.02+0.00}_{-0.00-0.02-0.00}$\\
				\cite{Zhou:2020ijj}	&$0.041$&$0.00295$&$0.0287$&$0.00173$\\
				\cite{Dash:2023hjr}	&$-$&$-$&$1.18$&$0.07$\\
				\hline\hline
			\end{tabular}\label{b3}}
	\end{center}
\end{table}
\begin{table}[H]
	\caption{The branching ratios $(10^{-3})$ of the decays $\bar B^0_s\to D^*_s(1S,2S)(\pi,\rho,K^{(*)})$, together with other theoretical results and data for comparison.}
	\begin{center}
		\scalebox{0.8}{
			\begin{tabular}{|c|c|c|c|c|c|c|c|c|c||}
				\hline\hline
				Modes&This work&\cite{Faustov:2012mt}&\cite{Blasi:1993fi}&\cite{Chen:2011ut}&\cite{KRF}&\cite{lirh}&\cite{Zhou:2020ijj}&\cite{Dash:2023hjr}&Exp.\cite{PDGT2024}\\
				\hline
				$\bar{B}_{s}^{0}\to D_{s}^{*+}(1S) \pi^{-}$&$4.03^{+0.02+1.30+0.59}_{-0.02-1.18-0.45}$&$2.7$&$2$&$3.1$&$2.11$&$2.42$&$3.37$&$-$&$1.9$\\
				$\bar{B}_{s}^{0}\to D_{s}^{*+}(1S) K^{-}$&$0.31^{+0.00+0.09+0.05}_{-0.00-0.09-0.03}$&$0.21$&$0.2$&$0.24$&$0.159$&$0.165$&$0.249$&$-$&$0.132$\\
				$\bar{B}_{s}^{0}\to D_{s}^{*+}(1S) \rho^{-}$&$11.19^{+0.04+3.34+1.55}_{-0.04-3.01-1.15}$&$8.7$&$13$&$9.0$&$-$&$5.69$&$7.26$&$11.73$&$9.5$\\
				$\bar{B}_{s}^{0}\to D_{s}^{*+}(1S) K^{*-}$&$0.65^{+0.00+0.19+0.09}_{-0.17-0.01-0.07}$&$0.48$&$0.6$&$0.56$&$0.163$&$0.347$&$0.688$&$0.69$&$-$\\
				\hline
				$\bar{B}_{s}^{0}\to D_{s}^{*+}(2S) \pi^{-}$&$0.61^{+0.00+0.12+0.07}_{-0.00-0.10-0.07}$&$0.8$&$-$&$-$&$-$&$-$&$0.108$&$-$&$-$\\
				$\bar{B}_{s}^{0}\to D_{s}^{*+}(2S) K^{-}$&$0.05^{+0.00+0.01+0.01}_{-0.00-0.01-0.01}$&$0.06$&$-$&$-$&$-$&$-$&$0.00777$&$-$&$-$\\
				$\bar{B}_{s}^{0}\to D_{s}^{*+}(2S) \rho^{-}$&$1.51^{+0.01+0.30+0.22}_{-0.01-0.26-0.22}$&$2.2$&$-$&$-$&$-$&$-$&$0.0475$&$1.05$&$-$\\
				$\bar{B}_{s}^{0}\to D_{s}^{*+}(2S) K^{*-}$&$0.09^{+0.00+0.02+0.01}_{-0.00-0.01-0.01}$&$0.12$&$-$&$-$&$-$&$-$&$0.00332$&$0.06$&$-$\\
				\hline\hline
			\end{tabular}\label{b4}}
	\end{center}
\end{table}

\begin{table}[H]
	\caption{The ratios of the decays $\bar B{(s)} \to D^{(*)+}_{(s)}(1S,2S)(\pi,\rho, K^{(*)})^{-}$. The results from the QCDF and the BS equation approaches and the available data are also listed for comparison.}
	\begin{center}
		\scalebox{1}{
			\begin{tabular}{|c|c|c|c|c|c|c|c|}
				\hline\hline
				Ratios&This work&$\cite{Zhou:2020ijj}$&\cite{Cai:2021mlt}&Exp.\cite{PDGT2024}&Ratios&This work&$\cite{Zhou:2020ijj}$ \\
				\hline\hline
				$\frac{Br(\bar{B}_{d} \to D^{*+}\pi^{-})}{Br(\bar{B}_{d} \to D^{+}\pi^{-})}$&$1.20$&$1.17$&$0.90$&$1.06$&$\frac{Br(\bar{B}_{d} \to D^{*+}(2S)\pi^{-})}{Br(\bar{B}_{d} \to D^{+}(2S)\pi^{-})}$&$0.95$&$0.83$\\
				$\frac{Br(\bar{B}_{d} \to D^{+}\rho^{-})}{Br(\bar{B}_{d} \to D^{+}\pi^{-})}$&$2.31$&$2.44$&$2.61$&$3.03$&$\frac{Br(\bar{B}_{d} \to D^{+}(2S)\rho^{-})}{Br(\bar{B}_{d} \to D^{+}(2S)\pi^{-})}$&$2.21$&$2.25$\\
				$\frac{Br(\bar{B}_{d} \to D^{+}\rho^{-})}{Br(\bar{B}_{d} \to D^{*+}\pi^{-})}$&$1.92$&$2.08$&$2.91$&$2.86$&$\frac{Br(\bar{B}_{d} \to D^{+}(2S)\rho^{-})}{Br(\bar{B}_{d} \to D^{*+}(2S)\pi^{-})}$&$2.34$&$2.71$\\
				\hline\hline
				$\frac{Br(\bar{B}_{d} \to D^{*+}K^{-})}{Br(\bar{B}_{d} \to D^{+}K^{-})}$&$1.18$&$1.63$&$0.89$&$1.05$&$\frac{Br(\bar{B}_{d} \to D^{*+}(2S)K^{-})}{Br(\bar{B}_{d} \to D^{+}(2S)K^{-})}$&$0.81$&$0.81$\\
				$\frac{Br(\bar{B}_{d} \to D^{+}K^{*-})}{Br(\bar{B}_{d} \to D^{+}K^{-})}$&$1.65$&$1.76$&$1.72$&$2.20$&$\frac{Br(\bar{B}_{d} \to D^{+}(2S)K^{*-})}{Br(\bar{B}_{d} \to D^{+}(2S)K^{-})}$&$1.57$&$1.64$\\
				$\frac{Br(\bar{B}_{d} \to D^{+}K^{*-})}{Br(\bar{B}_{d} \to D^{*+}K^{-})}$&$1.40$&$1.08$&$1.94$&$2.08$&$\frac{Br(\bar{B}_{d} \to D^{+}(2S)K^{*-})}{Br(\bar{B}_{d} \to D^{*+}(2S)K^{-})}$&$1.93$&$2.03$\\
				\hline\hline
				$\frac{Br(\bar{B}_{d} \to D^{+}K^{-})}{Br(\bar{B}_{d} \to D^{+}\pi^{-})}$&$0.078$&$0.076$&$0.076$&$0.082$&$\frac{Br(\bar{B}_{d} \to D^{+}(2S)K^{-})}{Br(\bar{B}_{d} \to D^{+}(2S)\pi^{-})}$&$0.078$&$0.074$\\
				$\frac{Br(\bar{B}_{d} \to D^{*+}K^{-})}{Br(\bar{B}_{d} \to D^{*+}\pi^{-})}$&$0.076$&$0.074$&$0.075$&$0.081$&$\frac{Br(\bar{B}_{d} \to D^{*+}(2S)K^{-})}{Br(\bar{B}_{d} \to D^{*+}(2S)\pi^{-})}$&$0.067$&$0.072$\\
				$\frac{Br(\bar{B}_{d} \to D^{+}K^{*-})}{Br(\bar{B}_{d} \to D^{+}\rho^{-})}$&$0.056$&$0.054$&$0.050$&$0.059$&$\frac{Br(\bar{B}_{d} \to D^{+}(2S)K^{*-})}{Br(\bar{B}_{d} \to D^{+}(2S)\rho^{-})}$&$0.055$&$0.054$\\
				\hline\hline
				$\frac{Br(\bar{B}_{s} \to D^{+}_{s}\pi^{-})}{Br(\bar{B}_{d} \to D^{+}K^{-})}$&$13.41$&$11.92$&$13.30$&$14.54$&$\frac{Br(\bar{B}_{s} \to D^{+}_{s}(2S)\pi^{-})}{Br(\bar{B}_{d} \to D^{+}(2S)K^{-})}$&$12.68$&$33.24$\\
				$\frac{Br(\bar{B}_{s} \to D^{+}_{s}\pi^{-})}{Br(\bar{B}_{d} \to D^{+}\pi^{-})}$&$1.05$&$0.90$&$1.01$&$1.19$&$\frac{Br(\bar{B}_{s} \to D^{+}_{s}(2S)\pi^{-})}{Br(\bar{B}_{d} \to D^{+}(2S)\pi^{-})}$&$0.99$&$2.47$\\
				\hline\hline
			\end{tabular}\label{ratios}	}
	\end{center}
\end{table}
To test the factorization hypothesis, as well as the SU(3) relations in $B$ meson decays, we can define the ratios of the branching fractions for our considered decays and list the values in Table \ref{ratios}, together with other theoretical results and experimental data.
From Table \ref{ratios}, one can find that our predictions are consistent with the results from the QCDF \cite{Huber:2016xod} and the BS equation \cite{Zhou:2020ijj} approaches and the available data \cite{PDGT}. The ratios $\frac{Br(\bar{B}_{s} \to D^{+}_{s}\pi^{-})}{Br(\bar{B}_{d} \to D^{+}K^{-})}$ and $\frac{Br(\bar{B}_{s} \to D^{+}_{s}\pi^{-})}{Br(\bar{B}_{d} \to D^{+}\pi^{-})}$ are used to determine the ratio of fragmentation functions $f_{d}/f_{s}$, which is an important quantity for precise
measurements of absolute $B_{s}$ decay rates in experiments \cite{Fleischer:2010ca,Fleischer:2010ay}. As for the case with the first radially excited charm meson involved, except for the ratios $\frac{Br(\bar{B}_{s} \to D^{+}_{s}(2S)\pi^{-})}{Br(\bar{B}_{d} \to D^{+}(2S)K^{-})}$ and $\frac{Br(\bar{B}_{s} \to D^{+}_{s}(2S)\pi^{-})}{Br(\bar{B}_{d} \to D^{+}(2S)\pi^{-})}$, which show relatively large differences, the results of other ratios agree well with the results given by the BS equation \cite{Zhou:2020ijj}. So we recommend future experiments measuring these two ratios to clarify which method is more reliable. 

\begin{table}[H]
	\caption{Polarization fractions $(\%)$ of the decays $B_{(s)}^{0}\to D^*_{(s)}(1S,2S)(\rho,K^*)$. $f_{L}$ and $ f_{\|}$ refer to the longitudinal and transverse parallel polarization fractions, respectively. In Ref. \cite{Chen:2011ut}, BS+FA refers that the form factors are obtained in the BS equation approach and the amplitudes are evaluated under the factorization approximation (FA). BS+PQCD refers that the nonfactorizable and annihilation contributions are further included under the PQCD approach. }
	\begin{center}
		\scalebox{1}{
			\begin{tabular}{|c|c|c|c|c|}
				\hline\hline
				Channel&$\bar{B}^{0}\to D^{*+}(1S) \rho^{-}$&$\bar{B}^{0}\to D^{*+}(1S) K^{*-}$&$\bar{B}^{0}\to D^{*+}(2S) \rho^{-}$&$\bar{B}^{0}\to D^{*+}(2S) K^{*-}$ \\
				\hline\hline
				$f_{L}[\%]$&$90.59$&$88.11$&$93.52$&$91.56$\\
                \hline                    
				$f_{\|}[\%]$&$8.27$&$10.47$&$6.42$&$8.36$\\
				\hline\hline
				Channel&$B^{-}\to D^{*0}(1S) \rho^{-}$&$B^{-}\to D^{*0}(1S) K^{*-}$&$B^{-}\to D^{*0}(2S) \rho^{-}$&$B^{-}\to D^{*0}(2S) K^{*-}$ \\
				\hline\hline
				$f_{L}[\%]$&$90.60$&$88.12$&$93.25$&$91.21$\\
                 \hline
				$f_{\|}[\%]$&$8.26$&$10.46$&$6.68$&$8.70$\\
				\hline\hline
				Channel&$\bar{B}_{s}^{0}\to D_{s}^{*+}(1S) \rho^{-}$&$\bar{B}_{s}^{0}\to D_{s}^{*+}(1S) K^{*-}$&$\bar{B}_{s}^{0}\to D_{s}^{*+}(2S) \rho^{-}$ &$\bar{B}_{s}^{0}\to D_{s}^{*+}(2S) K^{*-}$ \\
				\hline\hline
				$f_{L}[\%]$&$89.13$&$86.44$&$92.08$&$89.84$\\
				BS+FA$\cite{Chen:2011ut}$&$87.40$&$84.10$&$-$&$-$\\
				BS+PQCD$\cite{Chen:2011ut}$&$85.4$&$85.7$&$-$&$-$\\                
				PQCD$\cite{Li:2009xf}$&$87$&$83$&$-$&$-$\\
				$\cite{LRL}$&$105.00$&$-$&$-$&$-$\\
                \hline
				$f_{\|}[\%]$&$9.11$&$11.47$&$7.65$&$9.90$\\
				BS+FA \cite{Chen:2011ut}&$10.4$&$13.3$&$-$&$-$\\
                    BS+PQCD \cite{Chen:2011ut}&$11.3$&$10.4$&$-$&$-$\\
				\hline\hline
\end{tabular}\label{BR10}}
	\end{center}
\end{table}
 The polarization fractions for the decays $\bar{B}_{(s)}^{0}\to D^{*+}_{(s)}(1S,2S)(\rho, K^*)^-$ are listed in Table \ref{BR10}, where the results of the BS equation \cite{Chen:2011ut} and the PQCD \cite{Li:2009xf} approaches and the available data \cite{LRL} are also listed for comparison. From Table \ref{BR10}, it could be found that our predictions are comparable to the data and other theoretical results. 
Although there exist significant discrepancies between the form factors of the transitions $B_{(s)}\to D^*_{(s)}$ and $B_{(s)}\to D^*_{(s)}(2S)$, similar polarization behaviors can be observed in these decays 
$B_{(s)}\to D^{*}_{(s)}(1S,2S)(\rho, K^*)$, that is, the longitudinal polarization is dominant, reaching approximately $90\%$, while the transverse parallel and perpendicular polarization factions are only a few percent or roughly $10\%$. It shows that the form factors have little impact on polarization. 
\section{Summary}\label{sum}
In this article, we have provided a detailed study of the nonleptonic decays $B_{(s)}\to D^{(*)}_{(s)}(1S,2S)(\pi,\rho, K^{(*)})$ in the framework of the covariant light-front approach. Except for the predictions for neutral decays $B_{(s)}\to D^{(*)}_{(s)}(1S)(\pi,\rho, K^{(*)})$, which have some excess compared to the data, overall the branching ratios for the decays involving a ground state $D^{(*)}$ or  $D^{(*)}_s$ meson are consistent with the experimental measurements. As for the decays $B_{(s)}\to D^{(*)}_{(s)}(2S)(\pi,\rho, K^{(*)})$, most of their branching ratios lie in the range $10^{-5}\sim10^{-4}$, which are likely to be detected by the present LHCb and Belle II experiments. Our predictions for the decays with the first radially excited $D^{(*)}(2S)$ or $D^{(*)}_{s}(2S)$ meson involved are larger than the results given by the BS equation approach, but agree well with the RQM and RIQM calculations. We hope that future experiments can clarify these discrepancies. Although there exists an obvious difference in the branching ratios between the decays $B_{(s)}\to D^*_{(s)}(1S)(\rho,K^*)$ and $B_{(s)}\to D^*_{(s)}(2S)(\rho,K^*)$, similar polarization behaviors can be observed in them, that is, the longitudinal polarization is dominant, reaching approximately $90\%$, while the transverse parallel and perpendicular polarization fractions are only a few percent or roughly $10\%$. These studies are of great significance for our understanding of the first radially excited charmed mesons, even the spectrum of charmed mesons.     
\section*{Acknowledgment}
This work is partly supported by the National Natural Science
Foundation of China under grant No. 11347030 and the Natural Science Foundation of Henan
Province under grant No. 232300420116, 252300421302.
\appendix
\section{Some specific rules under the $p^-$ integration}
When performing the integration, we need to include the zero-mode contribution. It amounts to performing the integration in a proper way in the CLFQM. Specifically, we
use the following rules given in Refs. \cite{jaus,Y. Cheng}
\be
\hat{p}_{1 \mu}^{\prime} &\doteq &   P_{\mu}
A_{1}^{(1)}+q_{\mu} A_{2}^{(1)},\\
\hat{p}_{1 \mu}^{\prime}
\hat{p}_{1 \nu}^{\prime}  &\doteq & g_{\mu \nu} A_{1}^{(2)} +P_{\mu}
P_{\nu} A_{2}^{(2)}+\left(P_{\mu} q_{\nu}+q_{\mu} P_{\nu}\right)
A_{3}^{(2)}+q_{\mu} q_{\nu} A_{4}^{(2)},\\
Z_{2}&=&\hat{N}_{1}^{\prime}+m_{1}^{\prime 2}-m_{2}^{2}+\left(1-2
x_{1}\right) M^{\prime 2} +\left(q^{2}+q \cdot P\right)
\frac{p_{\perp}^{\prime} \cdot q_{\perp}}{q^{2}},\\
\hat{p}_{1 \mu}^{\prime} \hat{N}_{2} & \rightarrow & q_{\mu}\left[A_{2}^{(1)} Z_{2}+\frac{q \cdot P}{q^{2}} A_{1}^{(2)}\right],
\en
\be
\hat{p}_{1 \mu}^{\prime} \hat{p}_{1 \nu}^{\prime} \hat{N}_{2} & \rightarrow &g_{\mu \nu} A_{1}^{(2)} Z_{2}+q_{\mu} q_{\nu}\left[A_{4}^{(2)} Z_{2}+2 \frac{q \cdot P}{q^{2}} A_{2}^{(1)} A_{1}^{(2)}\right],\\
A_{1}^{(1)}&=&\frac{x_{1}}{2}, \quad A_{2}^{(1)}=
A_{1}^{(1)}-\frac{p_{\perp}^{\prime} \cdot q_{\perp}}{q^{2}},\quad A_{3}^{(2)}=A_{1}^{(1)} A_{2}^{(1)},\\
A_{4}^{(2)}&=&\left(A_{2}^{(1)}\right)^{2}-\frac{1}{q^{2}}A_{1}^{(2)},\quad A_{1}^{(2)}=-p_{\perp}^{\prime 2}-\frac{\left(p_{\perp}^{\prime}
\cdot q_{\perp}\right)^{2}}{q^{2}}, \quad A_{2}^{(2)}=\left(A_{1}^{(1)}\right)^{2}.  \en

\section{EXPRESSIONS OF $B_{(s)} \rightarrow P,V$ FORM FACTORS}
\be
S_{\mu \nu}^{P V}&=&\left(S_{V}^{P V}-S_{A}^{P V}\right)_{\mu \nu}\non
&=&\operatorname{Tr}\left[\left(\gamma_{\nu}-\frac{1}{W_{V}^{\prime \prime}}\left(p_{1}^{\prime \prime}-p_{2}\right)_{\nu}\right)\left(p_{1}^{\prime \prime}
+m_{1}^{\prime \prime}\right)\left(\gamma_{\mu}-\gamma_{\mu} \gamma_{5}\right)\left(\not p_{1}^{\prime}+m_{1}^{\prime}\right) \gamma_{5}\left(-\not p_{2}
+m_{2}\right)\right] \non
&=&-2 i \epsilon_{\mu \nu \alpha \beta}\left\{p_{1}^{\prime \alpha} P^{\beta}\left(m_{1}^{\prime \prime}-m_{1}^{\prime}\right)
+p_{1}^{\prime \alpha} q^{\beta}\left(m_{1}^{\prime \prime}+m_{1}^{\prime}-2 m_{2}\right)+q^{\alpha} P^{\beta} m_{1}^{\prime}\right\} \non
&&+\frac{1}{W_{V}^{\prime \prime}}\left(4 p_{1 \nu}^{\prime}-3 q_{\nu}-P_{\nu}\right) i \epsilon_{\mu \alpha \beta \rho} p_{1}^{\prime \alpha} q^{\beta} P^{\rho}\non &&
+2 g_{\mu \nu}\left\{m_{2}\left(q^{2}-N_{1}^{\prime}-N_{1}^{\prime \prime}-m_{1}^{\prime 2}-m_{1}^{\prime \prime 2}\right)
-m_{1}^{\prime}\left(M^{\prime \prime 2}-N_{1}^{\prime \prime}-N_{2}-m_{1}^{\prime \prime 2}-m_{2}^{2}\right)\right.\non
&&\left.-m_{1}^{\prime \prime}\left(M^{\prime 2}-N_{1}^{\prime}-N_{2}-m_{1}^{\prime 2}-m_{2}^{2}\right)
-2 m_{1}^{\prime} m_{1}^{\prime \prime} m_{2}\right\} \non &&
+8 p_{1 \mu}^{\prime} p_{1 \nu}^{\prime}\left(m_{2}-m_{1}^{\prime}\right)-2\left(P_{\mu} q_{\nu}
+q_{\mu} P_{\nu}+2 q_{\mu} q_{\nu}\right) m_{1}^{\prime}+2 p_{1 \mu}^{\prime} P_{\nu}\left(m_{1}^{\prime}-m_{1}^{\prime \prime}\right)\non &&
+2 p_{1 \mu}^{\prime} q_{\nu}\left(3 m_{1}^{\prime}-m_{1}^{\prime \prime}-2 m_{2}\right)
+2 P_{\mu} p_{1 \nu}^{\prime}\left(m_{1}^{\prime}+m_{1}^{\prime \prime}\right)+2 q_{\mu} p_{1 \nu}^{\prime}\left(3 m_{1}^{\prime}+m_{1}^{\prime \prime}-2 m_{2}\right)\non &&
+\frac{1}{2 W_{V}^{\prime \prime}}\left(4 p_{1 \nu}^{\prime}-3 q_{\nu}-P_{\nu}\right)\left\{2 p_{1 \mu}^{\prime}\left[M^{\prime 2}
+M^{\prime \prime 2}-q^{2}-2 N_{2}+2\left(m_{1}^{\prime}-m_{2}\right)\left(m_{1}^{\prime \prime}+m_{2}\right)\right]\right.\non&&
+q_{\mu}\left[q^{2}-2 M^{\prime 2}+N_{1}^{\prime}-N_{1}^{\prime \prime}+2 N_{2}-\left(m_{1}+m_{1}^{\prime \prime}\right)^{2}+2\left(m_{1}^{\prime}-m_{2}\right)^{2}\right]\non&&
\left.+P_{\mu}\left[q^{2}-N_{1}^{\prime}-N_{1}^{\prime \prime}-\left(m_{1}^{\prime}+m_{1}^{\prime \prime}\right)^{2}\right]\right\} .
\label{sptov}\en
\be
S_{\mu}^{P P}&=&\operatorname{Tr}\left[\gamma_{5}\left(\not p_{1}^{\prime \prime}+m_{1}^{\prime \prime}\right) \gamma_{\mu}\left(\not p_{1}^{\prime}
+m_{1}^{\prime}\right) \gamma_{5}\left(-\not p_{2}+m_{2}\right)\right] \non
&=& 2 p_{1 \mu}^{\prime}\left[M^{\prime 2}+M^{\prime \prime 2}-q^{2}-2 N_{2}-\left(m_{1}^{\prime}-m_{2}\right)^{2}-\left(m_{1}^{\prime \prime}
-m_{2}\right)^{2}+\left(m_{1}^{\prime}-m_{1}^{\prime \prime}\right)^{2}\right]\non
&&+q_{\mu}\left[q^{2}-2 M^{\prime 2}+N_{1}^{\prime}-N_{1}^{\prime \prime}+2 N_{2}+2\left(m_{1}^{\prime}-m_{2}\right)^{2}-\left(m_{1}^{\prime}
-m_{1}^{\prime \prime}\right)^{2}\right] \non
&&+P_{\mu}\left[q^{2}-N_{1}^{\prime}-N_{1}^{\prime \prime}-\left(m_{1}^{\prime}-m_{1}^{\prime \prime}\right)^{2}\right],
\label{ptop}
\en
The following are the analytical expressions of the $B_{(s)} \to D^{(\star)}_{(s)}(1S, 2S)$ transition form factors in the covariant light-front quark model
\begin{footnotesize}
\begin{eqnarray}
F^{B_{(s)} D_{(s)}}\left(q^{2}\right)&=&\frac{N_{c}}{16 \pi^{3}} \int d x_{2} d^{2} p_{\perp}^{\prime} \frac{h_{B_{(s)}}^{\prime}
h_{D_{(s)}}^{\prime \prime}}{x_{2} \hat{N}_{1}^{\prime} \hat{N}_{1}^{\prime \prime}}\left[x_{1}\left(M_{0}^{\prime 2}+M_{0}^{\prime \prime 2}\right)+x_{2} q^{2}\right.-x_{2}\left(m_{1}^{\prime}-m_{1}^{\prime \prime}\right)^{2}\non
&&-x_{1}\left(m_{1}^{\prime}-m_{2}\right)^{2}-x_{1}\left(m_{1}^{\prime \prime}-m_{2}\right)^{2}]\\
F^{B_{(s)} D_{(s)}}_{0}\left(q^{2}\right)&=&F^{B_{(s)} D_{(s)}}_{1}(q^{2})+\frac{q^2}{(q\cdot P)}\frac{N_{c}}{16 \pi^{3}}  \int d x_{2} d^{2} p_{\perp}^{\prime} \frac{2 h_{B_{(s)}}^{\prime}
h_{D_{(s)}}^{\prime \prime}}{x_{2} \hat{N}_{1}^{\prime} \hat{N}_{1}^{\prime \prime}}\left\{-x_{1} x_{2} M^{\prime 2}
-p_{\perp}^{\prime 2}-m_{1}^{\prime} m_{2}\right.\notag\\
&&+\left(m_{1}^{\prime \prime}-m_{2}\right)\left(x_{2} m_{1}^{\prime}
+x_{1} m_{2}\right)+2 \frac{q \cdot P}{q^{2}}\left(p_{\perp}^{\prime 2}+2 \frac{\left(p_{\perp}^{\prime} \cdot q_{\perp}\right)^{2}}{q^{2}}\right)
+2 \frac{\left(p_{\perp}^{\prime} \cdot q_{\perp}\right)^{2}}{q^{2}}\notag\\
&&\left.-\frac{p_{\perp}^{\prime} \cdot q_{\perp}}{q^{2}}\left[M^{\prime \prime 2}-x_{2}\left(q^{2}+q \cdot P\right)-\left(x_{2}-x_{1}\right) M^{\prime 2}+2 x_{1} M_{0}^{\prime 2}\right.\right.\notag\\
&&\left.\left.-2\left(m_{1}^{\prime}-m_{2}\right)\left(m_{1}^{\prime}+m_{1}^{\prime \prime}\right)\right]\right\},\\
V^{B_{(s)} D^{\star}_{(s)}}(q^{2})&=&\frac{N_{c}(M^{'}+M^{''})}{16 \pi^{3}} \int d x_{2} d^{2} p_{\perp}^{\prime} \frac{2 h_{B_{(s)}}^{\prime}
 h_{D^{\star}_{(s)}}^{\prime \prime}}{x_{2} \hat{N}_{1}^{\prime} \hat{N}_{1}^{\prime \prime}}\left\{x_{2} m_{1}^{\prime}
 +x_{1} m_{2}+\left(m_{1}^{\prime}-m_{1}^{\prime \prime}\right) \frac{p_{\perp}^{\prime} \cdot q_{\perp}}{q^{2}}\right.\non &&\left.
 +\frac{2}{w_{D^{\star}_{(s)}}^{\prime \prime}}\left[p_{\perp}^{\prime 2}+\frac{\left(p_{\perp}^{\prime} \cdot q_{\perp}\right)^{2}}{q^{2}}\right]\right\},\\
A_1^{B_{(s)} D^{\star}_{(s)}}(q^{2})&=& -\frac{1}{M^{'}+M^{''}}\frac{N_{c}}{16 \pi^{3}} \int d x_{2} d^{2} p_{\perp}^{\prime} \frac{h_{B_{(s)}}^{\prime} h_{D^{\star}_{(s)}}^{\prime \prime}}{x_{2}
\hat{N}_{1}^{\prime}
\hat{N}_{1}^{\prime \prime}}\left\{2 x_{1}\left(m_{2}-m_{1}^{\prime}\right)\left(M_{0}^{\prime 2}+M_{0}^{\prime \prime 2}\right)
-4 x_{1} m_{1}^{\prime \prime} M_{0}^{\prime 2}\right.\non
&&\left.+2 x_{2} m_{1}^{\prime} q \cdot P+2 m_{2} q^{2}-2 x_{1} m_{2}\left(M^{\prime 2}+M^{\prime \prime 2}\right)+2\left(m_{1}^{\prime}-m_{2}\right)\left(m_{1}^{\prime}
+m_{1}^{\prime \prime}\right)^{2}+8\left(m_{1}^{\prime}-m_{2}\right) \right.\non &&
\left. \times\left[p_{\perp}^{\prime 2}+\frac{\left(p_{\perp}^{\prime}
\cdot q_{\perp}\right)^{2}}{q^{2}}\right]+2\left(m_{1}^{\prime}+m_{1}^{\prime \prime}\right)\left(q^{2}+q \cdot P\right) \frac{p_{\perp}^{\prime} \cdot q_{\perp}}{q^{2}}
-4 \frac{q^{2} p_{\perp}^{\prime 2}+\left(p_{\perp}^{\prime} \cdot q_{\perp}\right)^{2}}{q^{2} w_{D^{\star}_{(s)}(nS)}^{\prime \prime}}
\right.\non && \left.\times\left[2 x_{1}\left(M^{\prime 2}+M_{0}^{\prime 2}\right)-q^{2}-q \cdot P-2\left(q^{2}+q \cdot P\right) \frac{p_{\perp}^{\prime} \cdot q_{\perp}}{q^{2}}-2\left(m_{1}^{\prime}-m_{1}^{\prime \prime}\right)\left(m_{1}^{\prime}-m_{2}\right)\right]\right\},\;\;\;\;\;\\
A_2^{B_{(s)} D^{\star}_{(s)}}(q^{2})&=& \frac{N_{c}(M^{'}+M^{''})}{16 \pi^{3}} \int d x_{2} d^{2} p_{\perp}^{\prime} \frac{2 h_{B_{(s)}}^{\prime} h_{D^{\star}_{(s)}}^{\prime \prime}}{x_{2} \hat{N}_{1}^{\prime}
\hat{N}_{1}^{\prime \prime}}\left\{\left(x_{1}-x_{2}\right)\left(x_{2} m_{1}^{\prime}+x_{1} m_{2}\right)-\frac{p_{\perp}^{\prime} \cdot q_{\perp}}{q^{2}}\left[2 x_{1} m_{2}
+m_{1}^{\prime \prime} \right.\right.\non &&
\left.\left.+\left(x_{2}-x_{1}\right) m_{1}^{\prime}\right]-2 \frac{x_{2} q^{2}+p_{\perp}^{\prime} \cdot q_{\perp}}{x_{2} q^{2} w_{D^{\star}_{(s)}}^{\prime \prime}}\left[p_{\perp}^{\prime} \cdot p_{\perp}^{\prime \prime}
+\left(x_{1} m_{2}+x_{2} m_{1}^{\prime}\right)\left(x_{1} m_{2}-x_{2} m_{1}^{\prime \prime}\right)\right]\right\},
\en
\be
A_0^{B_{(s)} D^{\star}_{(s)}}(q^{2})&=& \frac{M^{'}+M^{''}}{2M^{''}}A_1^{B_{(s)} D^{\star}_{(s)}}(q^{2})-\frac{M^{'}-M^{''}}{2M^{''}}A_2^{B_{(s)} D^{\star}_{(s)}}(q^{2})-\frac{q^2}{2M^{''}}\frac{N_{c}}{16 \pi^{3}} \int d x_{2} d^{2} p_{\perp}^{\prime} \frac{h_{B_{(s)}}^{\prime} h_{D^{\star}_{(s)}}^{\prime \prime}}{x_{2} \hat{N}_{1}^{\prime}
\hat{N}_{1}^{\prime \prime}}\non &&\left\{2\left(2 x_{1}-3\right)\left(x_{2} m_{1}^{\prime}+x_{1} m_{2}\right)-8\left(m_{1}^{\prime}-m_{2}\right)
\times\left[\frac{p_{\perp}^{\prime 2}}{q^{2}}
+2 \frac{\left(p_{\perp}^{\prime} \cdot q_{\perp}\right)^{2}}{q^{4}}\right]-\left[\left(14-12 x_{1}\right) m_{1}^{\prime}\right.\right. \non &&\left.\left.-2 m_{1}^{\prime \prime}-\left(8-12 x_{1}\right) m_{2}\right] \frac{p_{\perp}^{\prime} \cdot q_{\perp}}{q^{2}}
+\frac{4}{w_{D^{\star}_{(s)}}^{\prime \prime}}\left(\left[M^{\prime 2}+M^{\prime \prime 2}-q^{2}+2\left(m_{1}^{\prime}-m_{2}\right)\left(m_{1}^{\prime \prime}
+m_{2}\right)\right]\right.\right.\non &&\left.\left.\times\left(A_{3}^{(2)}+A_{4}^{(2)}-A_{2}^{(1)}\right)
+Z_{2}\left(3 A_{2}^{(1)}-2 A_{4}^{(2)}-1\right)+\frac{1}{2}\left[x_{1}\left(q^{2}+q \cdot P\right)
-2 M^{\prime 2}-2 p_{\perp}^{\prime} \cdot q_{\perp}\right.\right.\right.\non &&\left.\left.\left.-2 m_{1}^{\prime}\left(m_{1}^{\prime \prime}+m_{2}\right)
-2 m_{2}\left(m_{1}^{\prime}-m_{2}\right)\right]\left(A_{1}^{(1)}+A_{2}^{(1)}-1\right) q \cdot P\left[\frac{p_{\perp}^{\prime 2}}{q^{2}}
+\frac{\left(p_{\perp}^{\prime} \cdot q_{\perp}\right)^{2}}{q^{4}}\right]\right.\right.\non &&\left.\left.\times\left(4 A_{2}^{(1)}-3\right)\right)\right\}.\;\;\;
\end{eqnarray}
\end{footnotesize}

\end{document}